\documentclass[seceq,supplement]{ptptex}

\usepackage{graphicx}

\newcommand{\cobaltate}{\mbox{Na$_x$CoO$_2$}}
\newcommand{\vc}[1]{\boldsymbol{#1}}
\newcommand{\im}{\mathrm{i}}

\title{
Unusual electron correlations in Na$_{\mathbf{x}}$CoO$_{\mathbf{2}}$
due to the spin-state quasidegeneracy of cobalt ions
}

\author{
Ji\v{r}\'{\i} \textsc{Chaloupka}$^{1,2}$
and 
Giniyat \textsc{Khaliullin}$^1$
}

\inst{
$^1$Max-Planck-Institut f\"ur Festk\"orperforschung,
    Heisenbergstrasse 1, D-70569 Stuttgart, Germany \\
$^2$Department of Condensed Matter Physics, Faculty of Science,
    Masaryk University, Kotl\'a\v{r}sk\'a 2, 61137 Brno, Czech Republic
}

\abst{ 
Recent studies exposed many remarkable properties of layered cobaltates
\cobaltate{}. Surprisingly, many-body effects have been found to increase at
sodium-rich compositions of \cobaltate{} where one expects a simple, nearly
free motion of the dilute $S=1/2$ holes doped into a band insulator NaCoO$_2$.
Here we discuss the origin of enigmatic correlations that turn a doped
NaCoO$_2$ into a strongly correlated electronic system. A minimal model
including orbital degeneracy is proposed and its predictions are discussed.
The model is based on a key property of cobalt oxides -- the spin-state
quasidegeneracy of CoO$_6$ octahedral complex -- which has been known, {\it
e.g.}, in the context of an unusual physics of LaCoO$_3$ compound. Another
important ingredient of the model is the $90^\circ$ Co-O-Co bonding in
\cobaltate{} which allows nearest-neighbor $t_{2g}-e_g$ hopping. This hopping
introduces a dynamical mixture of electronic configurations $t_{2g}^6, S=0$
and $t_{2g}^5e_g^1, S=1$ of neighboring cobalt ions. We show that scattering
of charge carriers on spin-state fluctuations suppresses their coherent motion
and leads to the spin-polaron physics at $x\sim 1$.  At larger doping when
coherent fermionic bands are formed, the model predicts singlet
superconductivity of extended $s$-wave symmetry.  The presence of low-lying
spin states of Co$^{3+}$ is essential for the pairing mechanism. Implications
of the model for magnetic orderings are also discussed.
}

\begin{document}

\maketitle

\section{Introduction}
\label{sec:intro}

The physics of transition metal oxides offers a wealth of interesting
phenomena related to their strongly correlated nature. This is because the
bandwidth is relatively small compared to the intraionic Coulomb repulsion
between the $3d$ electrons. The understanding of many unique properties of
oxides, such as high-$T_c$ superconductivity and colossal magnetoresistivity
is therefore based on the well-known Mott physics.\cite{Mot74}

Among the various families of transition metal oxides, attention has recently
focused on layered cobaltates, in particular on the sodium cobaltate
Na$_x$CoO$_2$. The interest was initiated by the large thermoelectric power
observed in this compound,\cite{Ter97,Wan03,Lee06} \textit{i.e.} the 
capability of an efficient conversion of heat energy to electricity.
The research on these systems was further boosted by the unexpected discovery 
of superconductivity (SC) in water-intercalated \cobaltate{}.\cite{Tak03}
Soon after, many remarkable properties were found\cite{Ong04} such as
spin-sensitive thermopower,\cite{Wan03} unusual charge and spin
orderings,\cite{Foo04,Ber04,Bay05,Gas06,Ber07} very narrow quasiparticle
bands\cite{Qia06a,Shi06,Qia06b,Val02,Bro07} and especially a very unusual phase
diagram.\cite{Foo04} While the strongly correlated nature of \cobaltate{} is
no longer at doubt, the mechanisms by which the correlated electrons design
such an exotic phase diagram are not fully understood even on a qualitative
level.

\begin{figure}[tb]
\centerline{\includegraphics[width=9.6cm]{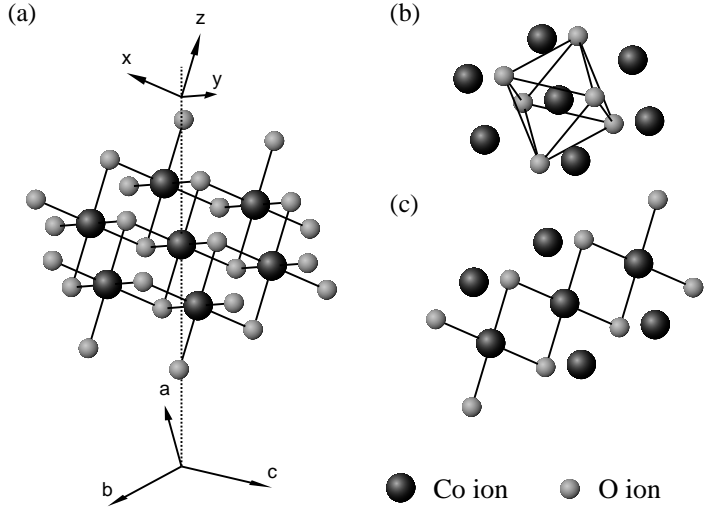}}
\caption{(a) Structure of a hexagonal CoO$_2$ plane with $90^\circ$ 
cobalt--oxygen bonds pointing along $x$, $y$ and $z$ directions.
The directions in the hexagonal lattice of cobalt ions are denoted by $a$, 
$b$ and $c$ respectively.
(b) Each cobalt ion is surrounded by six oxygen ions
forming an octahedron, in the real structure this is slightly distorted.
(c) Each cobalt ion is a member of three CoO$_2$ chains which determine planes
perpendicular to the $x$, $y$ and $z$ axes respectively. In the figure, chain 
of $b$ direction determining a plane perpendicular to the $y$ axis is shown.}
\label{fig:struct}
\end{figure}

Layered cobaltates consist of CoO$_2$ planes with a triangular lattice of
Co ions (see Fig.~\ref{fig:struct}), separated either by Na layers as in
\cobaltate{}\cite{Foo04} or by BiO-BaO layers of rock-salt structure in
so-called ``misfit'' cobaltates [see \citen{Bob07,Bro07} and references
therein]. By suppressing electron motion along the $c$-axis these layers impose
quasi-two-dimensionality upon the CoO$_2$ layers. In addition they provide the
charge carriers to the CoO$_2$ planes in doped compounds. Depending on the
composition, the valence state of Co ions may vary in a wide range from
non-magnetic Co$^{3+}$ $t_{2g}^6$ $S=0$ state (as in NaCoO$_2$) towards the
magnetic Co$^{4+}$ $t_{2g}^5$ $S=1/2$ configuration (in \cobaltate{} at small
$x$).

The Co$^{3+}$ systems where the cobalt ions have full $t_{2g}^6$ shell are
naturally referred to as band insulators,\cite{Vau05,Lan05} whereas in
Co$^{4+}$ $S=1/2$ rich systems strongly correlated Mott behavior is expected.
Thus by reducing sodium content in \cobaltate{} one should be able to observe
an evolution from the weakly correlated band-insulator regime to a doped Mott
limit. However, the experiments show completely opposite trends. The
Co$^{4+}$ rich compounds are just moderately correlated
metals\cite{Vau07,Foo04} while Co$^{3+}$ rich compositions show signatures of
strong correlations such as magnetic order,\cite{Foo04,Bay05} strong magnetic
field effects,\cite{Wan03} \textit{etc.} Pronounced incoherent structures in
the angular-resolved photoemission (ARPES) spectra\cite{Qia06,Bro07} 
revealing the complex structure of holes doped into NaCoO$_2$ are also 
enhanced near the band-insulator limit.

\begin{figure}[tb]
\centerline{\includegraphics[width=9.8cm]{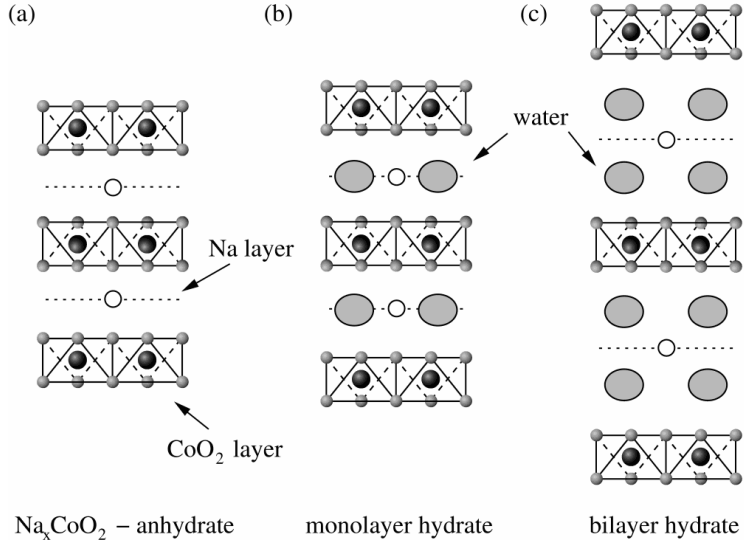}}
\caption{Layered structure of (a) anhydrate Na$_x$CoO$_2$ and 
water-intercalated (b) monolayer hydrate, (c) bilayer hydrate.
Water-intercalation leads to an expansion in the direction normal to CoO$_2$
planes and in the case (c) brings a large screening of sodium potential.}
\label{fig:layered}
\end{figure}

Even more puzzling situation occurs with hydration of \cobaltate{}. Depending
on the amount of water intercalating the structure, \cobaltate{} forms a
monolayer hydrate or (at a larger water content) a bilayer
hydrate\cite{Tak03a,Foo03} (see Fig.~\ref{fig:layered} for the details of the
structure). As a big surprise, superconductivity was discovered in the bilayer
hydrate of \cobaltate{} with $x\approx 0.35$.\cite{Tak03} The corresponding
transition temperature is quite low -- $T_c\simeq 5\:\mathrm{K}$ in the
optimal case. However, the identification of the pairing mechanism is a
problem of principal importance, because this may shed some light on the other
mysteries of \cobaltate{} as well. Soon after the discovery, comparisons with
the cuprate situation have been made.\cite{Tak03,Sch03} They emphasize the
quasi-two-dimensional CoO$_2$ layers of $S=1/2$ background (Co$^{4+}$ ions)
doped by $S=0$ (Co$^{3+}$ ions) charge carriers, thereby representing a
$t_{2g}$ analog of CuO$_2$ planes. The triangular lattice of Co ions
might be convenient for a realization of resonating-valence-bond (RVB)
state\cite{And87}, which was originally proposed for a $S=1/2$ Heisenberg
antiferromagnet on a triangular lattice.\cite{Faz74} Despite this, it was
quickly realized, that the phase diagram of \cobaltate{}\cite{Foo04} is
radically different from that of cuprates. As already mentioned, the enhanced
strongly correlated nature is rather observed at high relative content of
{\it nonmagnetic} Co$^{3+}$ ions. Also the location of the superconducting 
dome initially expected at cobalt valency around $3.65$ (as 
in \mbox{Na$_{0.35}$CoO$_2$}) was later found at actual valency $\sim 3.4$, 
closer to the Co$^{3+}$ situation due to water-induced valency 
shift.\cite{Tak04,Mil04,Kar04}

Many-body effects observed near a band insulator regime of cobaltates are
puzzling and indicate that electronic correlations originate from a specific
mechanism different from that of the cuprates. We have recently proposed a
model\cite{Kha08,Cha07} for strong correlations that operate over the entire
phase diagram of \cobaltate{} including a weakly doped band insulator limit.
In this paper, we discuss this model and its predictions in more detail, and
further extend it to the case of multiorbital situation.

Considering the limit of diluted $S=1/2$ Co$^{4+}$ holes doped into
non-magnetic band insulators NaCoO$_2$ and misfit cobaltates, the model
predicts a polaronic behavior of the holes characterized by strongly reduced
quasiparticle peaks and dispersive incoherent structures in their spectral
functions. In the Fermi-liquid regime at large concentration of the doped
holes, a specific pairing mechanism driven by correlated hopping of doped
holes emerges in the model. The resulting SC is optimized near the cobalt
valency $3.4$ in agreement with experimental observations. The model is based
on a unique aspect of Co$^{3+}$ ions, \textit{i.e.}, their spin-state
quasidegeneracy, and on special lattice geometry of CoO$_2$ layers realized in
the layered cobaltates.

\begin{figure}[tb]
\centerline{\includegraphics[width=12cm]{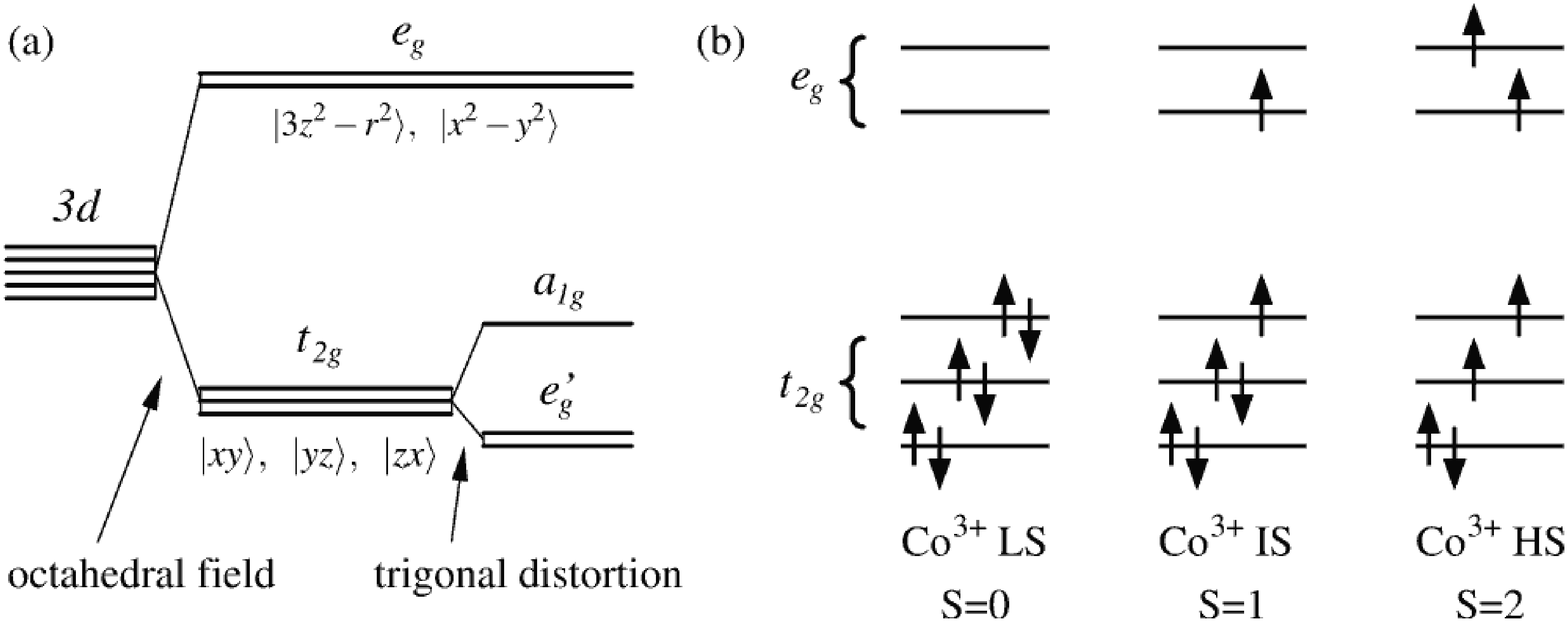}}
\caption{ (a) $3d$ levels of cobalt ions split into $t_{2g}$ and $e_g$ levels
in the octahedral field of oxygen ions. Trigonal distortion leads to a further
splitting among the $t_{2g}$ orbitals. The order of the resulting $a_{1g}$ and
$e'_g$ levels shown corresponds to that suggested by ARPES experiments in
\cobaltate{}. (b) Low-spin, intermediate-spin, and high-spin states of
Co$^{3+}$ ions. The competition between the $t_{2g}-e_g$ crystal field
splitting favoring $t_{2g}^6$ configuration and the Hund coupling favoring
higher spin values makes them quasidegenerate.}
\label{fig:levels}
\end{figure}

Let us first consider the $3d$ electrons of cobalt ions from a single electron
point of view as shown in Fig.~\ref{fig:levels}(a). The individual cobalt ions
are in octahedral crystal field [Fig.~\ref{fig:struct}(b)] which leads to the
usual $t_{2g}-e_g$ splitting $10Dq$ like in a perovskite structure. Due to the
layered structure of cobaltates, which imposes a quasi-two-dimensionality of
the electronic states, the symmetry of the originally three-fold degenerate
$t_{2g}$ levels is thus lowered and they split into so-called $a_{1g}$ and
$e'_g$ states. Crystal field prefers the $t_{2g}^5$ configuration on the
Co$^{4+}$ ions. The holes doped into NaCoO$_2$ or misfits are then expected to
occupy the highest $t_{2g}$ level. Initially, theoretical considerations have
suggested that this level should be of $e'_g$ symmetry (see Ref.~\citen{Pil08}
for details). However, experiments support an opposite scenario where the
doped holes prefer to occupy the $a_{1g}$ orbital states. Indeed, ARPES data
shows a very simple Fermi surface derived from a single band of $a_{1g}$
symmetry.\cite{Qia06,Bro07} 

In the case of Co$^{3+}$, the $t_{2g}^6$ $S=0$ configuration would be selected
by the cubic crystal-field splitting $10Dq$ alone. However, many-body effects
represented by the Hund coupling favoring high-spin configurations compete
with the crystal field splitting, making the $S=0$, $S=1$ and $S=2$ states of
Co$^{3+}$ quasidegenerate [Fig.~\ref{fig:levels}(b)]. Spin-state
quasidegeneracy of cobalt ions is well known, LaCoO$_3$ being a textbook
example.\cite{Mae04} According to Ref.~\citen{Hav06}, magnetic states are in
the range of $\sim 200-400\:\mathrm{meV}$ ($S=1$) and $\gtrsim
50\:\mathrm{meV}$ ($S=2$) above the $t_{2g}^6$ $S=0$ ground state (without
lattice relaxations). This leads to a distinct property of Co$^{3+}$ rich
oxides called ``spin-state-transition'' responsible for many anomalies such as
spin-state changes in LaCoO$_3$ \cite{Yam96,Hav06}, and a spin-blockade effect
in HoBaCo$_2$O$_{5.5}$ \cite{Mai04}, to mention a few manifestations of the
Janus-like behavior of Co$^{3+}$.

\begin{figure}[b]
\centerline{\includegraphics[width=13.5cm]{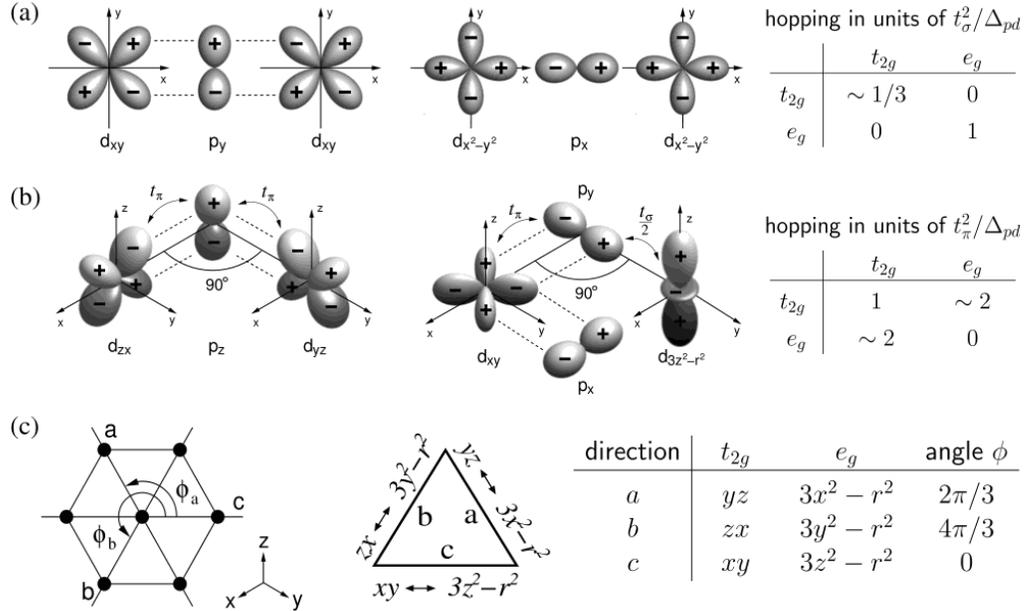}}
\caption{(a) Hopping between $d$-orbitals of cobalt ions via oxygen ions in
the case of $180^\circ$ Co-O-Co bonds. The geometry allows for the
nearest-neighbor $t_{2g}-t_{2g}$ and $e_g-e_g$ hopping. The $t_{2g}$ and $e_g$
sectors are independent. (b) The same for the $90^\circ$ Co-O-Co bonds as in
\cobaltate. Nearest-neighbor $t_{2g}-t_{2g}$ and $t_{2g}-e_g$ hopping is
possible, the latter one being approximately two times stronger.
(c) $a$, $b$ and $c$ directions in the hexagonal lattice of cobalt ions
and orbitals connected by the $t_{2g}-e_g$ hopping along the three
directions.}
\label{fig:hopgeom}
\end{figure}

In the groundstate the cobalt ions are supposed to be in $S=1/2$ Co$^{4+}$ or
$S=0$ Co$^{3+}$ states. Thanks to the spin-state quasidegeneracy, a new, third
degree of freedom -- namely the low-lying $S=1$ $t_{2g}^5e_g^1$ state of
Co$^{3+}$ -- could be in principle employed by intersite $t_{2g}-e_g$ hopping
which mixes the $t_{2g}^6$ and $t_{2g}^5e_g^1$ configurations of neighboring
ions.\footnote{The $S=2$ Co$^{3+}$ configuration with two $e_g$ electrons is
not accessible by the $t_{2g}-e_g$ hopping.} Whether such a matrix element is
finite or not depends on lattice geometry. With the $180^\circ$ Co--O--Co
bond geometry of perovskite cobaltates, the $t_{2g}$ and $e_g$ sectors are
separated with respect to the nearest neighbor hopping [see
Fig.~\ref{fig:hopgeom}(a) for an explanation], which is therefore incapable to
produce the $S=1$ Co$^{3+}$ states. New situation encountered in layered
cobaltates is that CoO$_6$ octahedra are edge-shared. In this geometry, the
hopping occurs along the $90^\circ$ Co--O--Co bonds, where the largest matrix
element is that between the orbitals of $t_{2g}$ and $e_g$ symmetry as shown
in Fig.~\ref{fig:hopgeom}(b). Doped holes can now easily generate $S=1$
$t_{2g}^5e_g^1$ states of Co$^{3+}$ which become strongly coupled to the
groundstate by virtue of intersite $t_{2g}-e_g$ electron transfer. In other
words, the magnetic configuration of Co$^{3+}$ is activated once the mobile
Co$^{4+}$ holes are added in \cobaltate{}. A dynamical generation of
$t_{2g}^5e_g^1$ $S=1$ states by a hole motion converts it into a many-body
correlated object -- the spin-polaron. At larger density of Co$^{4+}$ when
spin-polarons overlap, the virtual $S=1$ states act as mediators of an
effective spin-sensitive interaction. We eliminate these virtual states
perturbatively, and find an effective model in a form of spin-selective pair
hopping of electrons. The correlated hopping energy is optimized when holes
are paired and condense into a SC state. 

A r\'{e}sum\'{e} is that low-lying magnetic states of Co$^{3+}$, accessible
for electrons via the intersite hopping, provide an extra dimension in
physics of \cobaltate{}. In Sec.~\ref{sec:Htttilde}, we design a model
incorporating this idea. Based on this model, we demonstrate in
Sec.~\ref{sec:polaron} that holes doped into the band insulator NaCoO$_2$
behave in fact as magnetic polarons dressed by the spin-state fluctuations of
Co$^{3+}$ ions that are excited by the hole motion. Sec.~\ref{sec:HtP} derives
the effective interaction between holes, mediated by virtual spin-state
excitations of Co$^{3+}$ ions, in a Fermi-liquid regime at larger hole
densities. In Sec.~\ref{sec:spinsusc}, we discuss the relevance of these
interactions to the spin ordering, and, by exploring their effect on the spin
susceptibility, we find signatures of $2k_F$-instabilities. Finally, we focus
in Sec.~\ref{sec:SCsimple} and \ref{sec:SCfull} on the superconductivity
within our model. First, in Sec.~\ref{sec:SCsimple} we discuss symmetry and
doping dependencies of pairing instabilities in the model based on the
$a_{1g}$ band only. To address a possible role of $e'_g$ pockets in the
superconductivity, in Sec.~\ref{sec:SCfull} we extend the model by employing
the full orbital structure of the relevant states. Sec.~\ref{sec:conclusions}
concludes the paper.

\section{Model Hamiltonian}
\label{sec:Htttilde}

\begin{figure}[bt]
\centerline{\includegraphics[width=8cm]{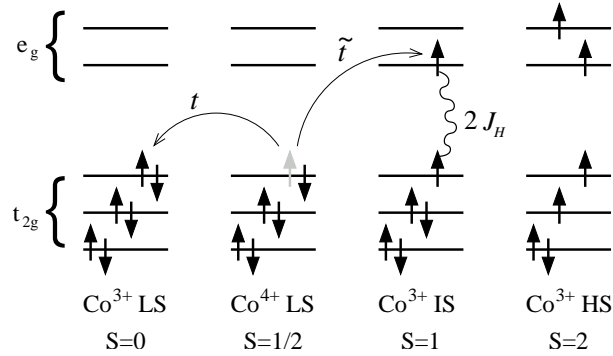}}
\caption{Processes contained in the $t-\tilde{t}$ model. Usual $t$-hopping
gives rise to the band structure. $\tilde{t}$-hopping allows to employ the
low-lying triplet state of Co$^{3+}$. High-spin ($S=2$) state is not
accessible by hopping.}
\label{fig:ttproc}
\end{figure}

The $t_{2g}$ orbitals in \cobaltate{} split into single
$a_{1g}=(d_{xy}+d_{yz}+d_{zx})/\sqrt{3}$ state and two $e'_g$ states:
$e'_{g1}=(d_{yz}-d_{zx})/\sqrt{2}$ and
$e'_{g2}=(2d_{xy}-d_{yz}-d_{zx})/\sqrt{6}$. The photoemission experiments
\cite{Qia06a,Shi06,Qia06b,Bro07} show that a single band, derived mostly from
the $a_{1g}$ orbitals, is active near the Fermi level [see Ref.~\citen{Zho05}
for possible orbital-selection mechanism]. Therefore, we consider first a
model based on the $a_{1g}\equiv f$ hole states (the version containing also
$e'_g$ orbitals will be introduced in Sec.~\ref{sec:SCfull}). Valence
fluctuations $d_j^6d_i^5\rightarrow d_j^5d_i^6$ (see the left part of
Fig.~\ref{fig:ttproc}) within the low-spin $t_{2g}$ manifold are described by
the tight-binding Hamiltonian 
\begin{equation}\label{eq:Ht}
H_t=-t\sum_{ij\sigma} f^\dagger_{j\sigma}f^{\phantom{\dagger}}_{i\sigma}\;,
\end{equation}
where $t=2t_0/3$ and $t_0=t_{\pi}t_{\pi}/\Delta_{pd}$ is the overlap between
$t_{2g}$ orbitals\cite{Kos03} (hereafter, a hole representation is used).
The fermionic operators $f$ are subject to a conventional Gutzwiller
constraint (at most one hole at a given site). As a result the bandwidth
following from \eqref{eq:Ht} is reduced by the Gutzwiller factor
$g_t=2n_d/(1+n_d)$ ($\approx n_d$ at the relevant dopings). Here $n_d$ is the
relative fraction of Co$^{3+}$ ions.

Our crucial observation following from Fig.~\ref{fig:hopgeom}(b) is that the
$t_{2g}$--$e_g$ hopping $\tilde t=t_{\sigma}t_{\pi}/\Delta_{pd}$, which uses a
stronger $\sigma$-bonding path with $t_{\sigma}/t_{\pi}\sim 2$, leads to more
effective valence fluctuations. These will constitute the second part,
$H_{\tilde{t}}$, of our model Hamiltonian:
$H_{t-\tilde{t}}=H_t+H_{\tilde{t}}$. The $\tilde t$ process
(Fig.~\ref{fig:ttproc}) generates $S=1$ state of Co$^{3+}$ composed of a
$t_{2g}$ hole and an $e_g$ electron. Low-spin $S=0$ $t_{2g}^5e_g^1$ state is
much higher in energy and can be ignored (see Sec.~\ref{sec:SCfull} for
details on the $t_{2g}^5e_g^1$ multiplet structure of Co ions). We represent
the resulting state by the operator ${\cal T}$ specified by the spin
projection of the $S=1$ state and the $e_g$ orbital $\gamma$ created by
$\tilde t$ hopping, {\it i.e.}, 
${\cal T}^\dagger_{+1,\gamma}=
 \overline{e^\dagger_{\gamma\uparrow}f^\dagger_\uparrow}$,
${\cal T}^\dagger_{-1,\gamma}=
 \overline{e^\dagger_{\gamma\downarrow}f^\dagger_\downarrow}$ and
${\cal T}^\dagger_{0,\gamma}=(\overline{
 e^\dagger_{\gamma\uparrow}f^\dagger_\downarrow+
 e^\dagger_{\gamma\downarrow}f^\dagger_\uparrow
})/\sqrt{2}$. 
The $e_g$ orbital is selected by the hopping geometry depicted in
Fig.~\ref{fig:hopgeom}(b). The nearest neighbor Co ions and two O ions binding
them determine a plane [see Fig.~\ref{fig:struct}(c)] which is labeled as $a$,
$b$ or $c$ according to the Co-Co bond direction. With respect to this plane,
the $\tilde{t}$-hopping couples the in-plane $t_{2g}$ orbital to the
out-of-plane $e_g$ orbital. This rule is shown in Fig.~\ref{fig:hopgeom}(c).
For the $\tilde{t}$ hopping along $a$, $b$, or $c$ bond, $\gamma=3x^2-r^2$,
$3y^2-r^2$ or $3z^2-r^2$ respectively. The $H_{\tilde{t}}$ part of our minimal
model Hamiltonian for \cobaltate{} reads then as\cite{Kha08} 
\begin{equation}
\label{eq:Htilde}
H_{\tilde t} = -\frac{\tilde t}{\sqrt 3} \sum_{ij} \Bigl[
{\cal T}^\dagger_{+1,\gamma}(i)
f^\dagger_{j\downarrow}f^{\phantom{\dagger}}_{i\uparrow}-
{\cal T}^\dagger_{-1,\gamma}(i)
f^\dagger_{j\uparrow}f^{\phantom{\dagger}}_{i\downarrow}
-{\cal T}^\dagger_{0,\gamma}(i)\,\tfrac1{\sqrt{2}}\!\left(
  f^\dagger_{j\uparrow}f^{\phantom{\dagger}}_{i\uparrow}-
  f^\dagger_{j\downarrow}f^{\phantom{\dagger}}_{i\downarrow}
\right)
+\mathrm{h.c.}\Bigr]\;.
\end{equation}
The factor of $1/\sqrt{3}$ comes from the projection onto the $a_{1g}$ band
within $t_{2g}$ sector. $H_{\tilde t}$ moves an electron from Co$_j^{3+}$ to
Co$_i^{4+}$ -- producing a $t_{2g}$ hole on site $j$ -- and replaces the
$t_{2g}$ hole on site $i$ by a complex excitation ${\cal T}$. Making use of
the $t_{2g}$--$e_g$ hopping (the {\it largest} one for 90$^{\circ}$ Co-O-Co
bonds), an electron ``picks-up'' the spin correlations present in virtual
$S=1$ states. 

The index $\gamma$ of the $\mathcal{T}$ excitation is determined by the
orientation of the $\langle ij\rangle$ bond according to the rules in
Fig.~\ref{fig:hopgeom}(c). The overlap between $e_g$ orbitals specified by
$\gamma$ and $\gamma'$ is
$\langle\gamma|\gamma'\rangle=\cos(\phi_\gamma-\phi_{\gamma'})$.
Consequently, the excitations ${\cal T}_\gamma$ inherit the same overlap: 
$\langle {\cal T}^{\phantom{\dagger}}_\gamma {\cal T}^\dagger_{\gamma'}\rangle
\propto \langle\gamma|\gamma'\rangle$. 
This brings peculiar geometrical factors into the theory, as will be observed
in the following sections.

The ${\cal T}$-excitation energy $E_T$ is determined by all the many-body
interactions within the CoO$_6$ complex (Hund's coupling, $p-d$ covalency,
crystal field, \ldots)\cite{Hav06}. This is a free parameter of the model.
Experimentally, $S=1$ states of CoO$_6$ complex in perovskite compound
LaCoO$_3$ are found at energies $E_T \sim 0.2-0.4\:\mathrm{eV}$ \cite{Hav06}
as already mentioned in the introduction. Based on this observation, we will
use in this paper the representative value $E_T \simeq 0.2-0.3\:\mathrm{eV}$
for layered cobaltates. In units of the $a_{1g}$ hopping integral $t\simeq
0.1\:\mathrm{eV}$ (which follows from the band structure fit $t_0\simeq
0.15\:\mathrm{eV}$ \cite{Zho05}), this translates into $E_T/t=2-3$ adopted
below in our numerical data. In principle, we expect some material dependence
of $E_T$ as it is decided by the balance of several competing interactions. It
is therefore highly desirable to quantify a multiplet structure of the
CoO$_6$-complex in \cobaltate{} as done in LaCoO$_3$ \cite{Hav06}. For the
ratio of the hopping amplitudes $\tilde{t}$ and $t_0$, we set $\tilde t/t_0=2$
as $t_{\sigma}/t_{\pi}\sim 2$.

\section{Spin-state polaron behavior of quasiparticles}
\label{sec:polaron}

Based on the model introduced in the previous section, we develop now a theory
for the photoemission spectra in cobaltates at large sodium content. The
concentration of the Co$^{4+}$ holes doped into originally non-magnetic band
insulator NaCoO$_2$ is small, so that we consider here the individual motion
of the holes only. It is evident from the Hamiltonian \eqref{eq:Htilde}, that
by creating and destroying ${\cal T}$ excitations as they propagate, the holes
are strongly renormalized and we deal with a spin-polaron problem. This
resembles the problem of doped Mott insulators like cuprates, however, the
nature of spin excitations is different here because of the nonmagnetic ground
state. Instead of magnon-like propagating modes as in cuprates, fluctuations
of the very spin value of Co$^{3+}$ ions are the cause of the spin-polaron
physics in cobaltates. In contrast to Refs.~\citen{Kha05,Dag06}, where a {\it
static} hole surrounded by $S=1$ Co$^{3+}$ ions was studied, in the present
model the triplet $S=1$ excitations are {\it virtual} and generated
dynamically by the {\it very motion} of the hole via the $\tilde{t}$ process.
These two pictures can merge if a hole is strongly trapped ({\it e.g.}, by
Na-potential\cite{Rog07}). 

\begin{figure}[tb]
\centerline{\includegraphics[width=8cm]{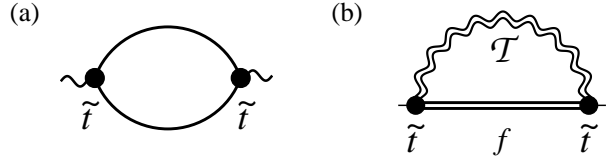}}
\caption{(a) Selfenergy of the $\mathcal{T}$ excitation in the lowest order 
approximation. Solid lines represent bare hole propagator. (b) Selfenergy of
the holes in a self-consistent Born approximation. Double straight line is 
the full hole propagator. Double wiggly line represents the propagator 
of the $\mathcal{T}$ excitation renormalized by the selfenergy in (a). 
(After Ref.~\citen{Cha07}.)}
\label{fig:diag}
\end{figure}

For the calculation of the fermionic self-energies, we employ the
selfconsistent Born approximation, which has extensively been used in the
context of spin-polarons in cuprates\cite{Kan89}. First, we focus on spin
excitation spectrum. Since a direct $e_g$-$e_g$ hopping in case of
$90^{\circ}$-bonds is not allowed by symmetry, the bare ${\cal T}$ spin
excitation is a purely local mode, at the energy $E_T$. The coupling to the
holes in \eqref{eq:Htilde} shifts and broadens this level. Accounting for this
effect perturbatively [see Fig.~\ref{fig:diag}(a) for a diagrammatic
representation], we obtain the ${\cal T}$ Green's function 
${\cal D}^{-1}(\im\omega)=\im\omega-E_T-\Sigma_T(\im\omega)$ with 
\begin{equation}
\label{tripselfE}
\Sigma_T(\im\omega)=\frac{2\tilde{t}^{\,2}}{3\beta}
\sum_{\vc k\vc k',\im\epsilon} \Gamma_{\vc k}\,
{\cal G}_0(\vc k,\im\epsilon) {\cal G}_0(\vc k',\im\epsilon+\im\omega) \;.
\end{equation}
Here, ${\cal G}_0$ is the bare electron propagator ${\cal G}_0(\vc
k,\im\epsilon)=(\im\epsilon -\xi_{\vc k})^{-1}$ with the $a_{1g}$ dispersion
on a triangular lattice $\xi_{\vc k}=-2t(c_a+c_b+c_c)+\mu$, where
$c_\gamma=\cos k_\gamma$ and $k_\gamma$ are the projections of $\vc k$ on the
directions $a$, $b$, $c$. The underlying $E_g$ symmetry of ${\cal T}$
operators involved in $\tilde{t}$ hopping results in the factor $\Gamma_{\vc
k}=c_a^2+c_b^2+c_c^2-c_ac_b-c_bc_c-c_cc_a$. We neglected a weak momentum
dependence of $\Sigma_T$ for the sake of simplicity. This is justified as long
as $\Sigma_T$ is small compared to the spin gap $E_T$. 

Further, we approximate $\Gamma_{\vc k}$ by its Brillouin-zone average $3/2$,
obtaining the simple expressions for $\Sigma_T$ in terms of bare fermionic
density of states $N_0(x)=\sum_{\vc k} \delta(x-\xi_{\vc k})$: 
\begin{equation}
\label{ImET}
\mathrm{Im}\Sigma_T(E)=
 -\pi\tilde{t}^{\,2}\int_{-E}^0 \mathrm{d}x\, N_0(x)N_0(x+E) \;,
\end{equation}
\begin{equation}
\label{ReET}
\mathrm{Re}\Sigma_T(E)=
 -\tilde{t}^{\,2}\int_{-\infty}^0\mathrm{d}x
\int_{x^2}^\infty \mathrm{d}y^2\,
\frac{N_0(x)N_0(x+y)}{y^2-E^2} \;.
\end{equation}
These equations determine the renormalized spectral function 
of the spin-triplet excitation:
$\rho_T(E) = -\pi^{-1} \mathrm{Im}
{\cal D}(\im\omega \rightarrow E+\im\delta)$.
Next, we use $\rho_T(E)$ below for calculation of the fermionic selfenergy. 

Holes are renormalized by creating and destroying $S=1$ Co$^{3+}$ excitations
while moving in the predominantly Co$^{3+}$ background.
The selfenergy diagram in a selfconsistent Born approximation accounting for
this effect [Fig.~\ref{fig:diag}(b)] reads as:
\begin{equation}\label{holeselfE}
\Sigma_{\vc k}(\im\epsilon)=-\frac{2\tilde{t}^{\,2}}{\beta}
\sum_{\vc k', \im\omega}\left[
\Gamma_{\vc k'} {\cal D}(-\im\omega)+
\Gamma_{\vc k} {\cal D}(\im\omega)\right]
{\cal G}(\vc k',\im\epsilon+\im\omega) \;,
\end{equation}
where ${\cal G}^{-1}(\vc k,\im\epsilon)= 
\im\epsilon-\xi_{\vc k}-\Sigma_{\vc k}(\im\epsilon)$. 
We can decompose the $\vc k$-dependence of the selfenergy into two simple
terms and write 
$\Sigma_{\vc k}(\omega)=
2\tilde{t}^{\,2}[\Phi(\omega)+\Gamma_{\vc k}\Xi(\omega)]$, 
where 
\begin{equation}\label{Phi}
\Phi(\omega)=\int_0^\infty\mathrm{d}E\, \rho_T(E)
\int_0^\infty \mathrm{d}x\, \frac{\tilde{N}(x)}{\omega-E-x+\im\delta} \;,
\end{equation}
\begin{equation}\label{Xi}
\Xi(\omega)=\int_0^\infty\mathrm{d}E\, \rho_T(E)
\int_{-\infty}^0 \mathrm{d}x\, \frac{N(x)}{\omega+E-x+\im\delta} \;.
\end{equation}
The full local density of states $N(E)=\sum_{\vc k} A({\vc k},E)$ and 
its $E_g$ symmetry part 
$\tilde{N}(E)=\sum_{\vc k} \Gamma_{\vc k} A({\vc k},E)$ are functions of
the selfenergy itself, via the spectral functions 
$A({\vc k},E)=-\pi^{-1}\mathrm{Im}\,{\cal G}(\im\epsilon\rightarrow
E+\im\delta)$. 
The above equations are thus to be solved self-consistently. 

The reliability of the approximations made was tested by the comparison with
an exact diagonalization of $H_{t-\tilde t}$. Considering a single hole
injected on a $7$-site hexagonal cluster, it was shown in Ref.~\citen{Cha07},
that the above equations give results consistent with the exact
diagonalization, even at rather small spin gap values $E_T\sim t$.

\begin{figure}[tb]
\centerline{\includegraphics[width=14cm]{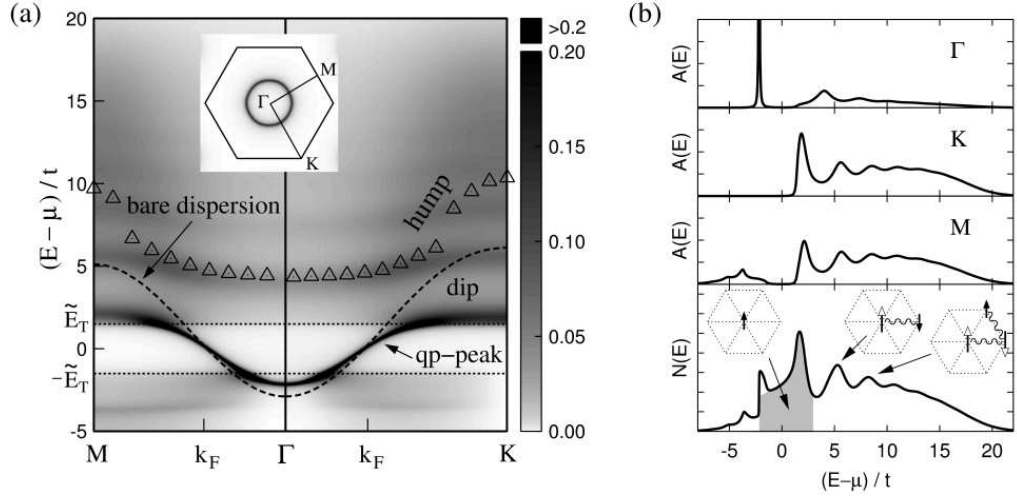}}
\caption{(a) Intensity map of the spectral function $A(\vc k, E)$ of the
Co$^{4+}$ holes along the $M-\Gamma-K$ path in the hexagonal Brillouin zone
(inset) calculated at 30\% doping and $E_T=2t$. Near the Fermi level, the
quasiparticle (qp-) peak is well developed. As the hole energy reaches the
renormalized $\mathcal{T}$ excitation energy $\tilde{E}_T$, the quasiparticle
peak broadens and its weight is transferred to a broad, incoherent background.
This results in a ``peak-dip-hump'' profile of the spectral function as seen
in the right panel. The top of the smoothed hump structure is indicated by
triangles. The dashed line shows the bare dispersion. (b) Spectral function
profiles at the $\Gamma$, $K$ and $M$ points and the total density of states
$N(E)$. The incoherent structure dominates $N(E)$, a small coherent part
resembling the density of states of the bare band (with a reduced bandwidth)
is indicated by the gray area. Several maxima on the hump structure reflect
the presence of multiple $\mathcal{T}$ excitations as sketched in the cartoon
figures. (After Ref.~\citen{Cha07}.)}
\label{fig:spect}
\end{figure}

To illustrate the gross features of the hole renormalization, in
Fig.~\ref{fig:spect} we show a complete map of the spectral function along
$M-\Gamma-K$ path in the Brillouin zone and spectral profiles at selected
points. We have used a representative value $E_T=2t$ which is renormalized by
holes to $\tilde{E}_T\approx 1.4t$. Renormalization of the holes leads to
spectral functions with a reduced quasiparticle (qp-) peak whose spectral
weight is transferred to a pronounced hump structure. [A peculiar momentum
dependence of the matrix elements $\Gamma_{\vc k}$ (note that $\Gamma_{\vc
k=0}=0$) reduces the effect at $\vc k=\Gamma$ point]. Compared to the bare
dispersion, the bandwidth of the renormalized holes is reduced by a factor
$\sim 2$. The main observation here is that as the hole energy reaches
$\tilde{E}_T$, the dynamical generation of $S=1$ excitations becomes very
intense and a broad incoherent response develops, leading to the pronounced
``peak-dip-hump'' structure of $A({\vc k},E)$. Several maxima on the hump
reflect the presence of multiple triplet excitations created by the hole
propagation. All these are the typical signatures of polaron physics.
Multiplet structure of the hump will in reality be smeared by phonons which
are naturally coupled to the $\tilde t$ transition involving also the orbital
sector. Although experiments\cite{Rue06} indicate that electron-phonon
coupling is moderate in cobaltates, it may enhance the spin-polaron effects as
in cuprates.\cite{Mis04} Following the maximum of the smoothed hump
structure, we observe its strong dispersion (stemming also from incoherent
$\tilde t$ hopping). 

\begin{figure}[tb]
\centerline{\includegraphics[width=7.5cm]{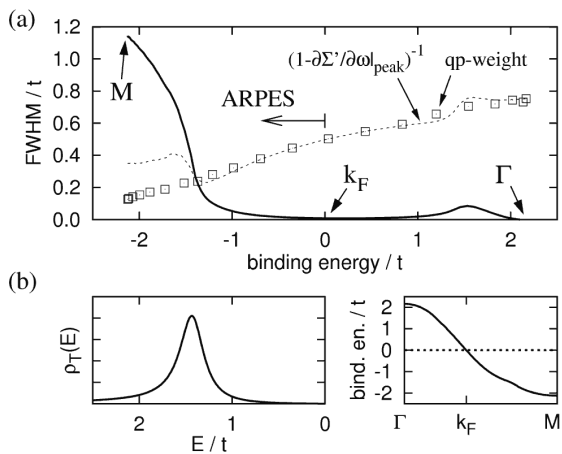}}
\caption{Top panel: Full energy width at half maximum of the quasiparticle
peak along the $M-\Gamma$ dispersion curve (see bottom-right panel) plotted as
a function of the binding energy. The part from $k_F$ to $M$ is accessible by
ARPES experiments. Strong quasiparticle damping below $\sim -1.4t$ is due to
a scattering on $S=1$ excitations. The spectral weight of the quasiparticle
peak obtained by a direct integration is indicated by squares. When damping is
small, it coincides with a conventional quasiparticle residue
$(1-\partial\Sigma'/\partial\omega)^{-1}$. Lower panel: The ${\cal T}$-exciton
spectral function (left), and renormalized hole dispersion (right). 
(After Ref.~\citen{Cha07}.)}
\label{fig:spectFWHM}
\end{figure}

Fig.~\ref{fig:spect} suggests a possible determination of $\tilde{E}_T$ from
the quasiparticle damping. To address this problem, in
Fig.~\ref{fig:spectFWHM} we show the energy width of the qp-peak following its
dispersion curve. The sharp onset of the damping at the binding energy
$\approx -1.4t$ is clearly related to the maximum of the spin-excitation
spectral function $\rho_T(E)$. In addition, Fig.~\ref{fig:spectFWHM} shows
the weight of the qp-peak which is $\vc k$-dependent (mainly due to the matrix
element $\Gamma_{\vc k}$).

Comparison of Figs.~\ref{fig:spect}, \ref{fig:spectFWHM} with the data of
Refs.~\citen{Bro07,Qia06} reveals a remarkable correspondence between theory
and experiment. In particular, both the qp-peak and the hump dispersions (see
Figs.~2, 3 of Ref.~\citen{Bro07}) are well reproduced by theory, considering
$t\approx 0.1\:\mathrm{eV}$ suggested by the band structure fit.\cite{Zho05}
The onset energy $\tilde{E}_T\sim 1.4t$ for the qp-damping
(Fig.~\ref{fig:spectFWHM}) is then $\approx 0.14\:\mathrm{eV}$, in agreement
with experiment (see Fig.~2(c) of Ref.~\citen{Bro07}).

Physically, at large $x\sim 1$ limit, dilute polarons are readily trapped by a
random potential of Na-vacancies\cite{Rog07,Mar07}, thus qp-peaks should be
suppressed at low hole doping. When the binding is strong, physics is local
and a polaron takes a form of hexagon-shaped $S=1/2$ object where a hole is
oscillating to optimize both $t$ and $\tilde t$ channels\cite{Kha05}. Our
model provides a microscopic basis for spin-polarons introduced on
experimental grounds\cite{Ber04,Ber07} and discussed in detail in
Refs.~\citen{Kha05,Dag06}. When the density of polarons is increased (as $x$
decreases), they start to overlap forming narrow bands. Eventually, the
polaron picture breaks down and a correlated Fermi-liquid emerges when $x$ is
further reduced. This is the subject of the following section, where we
develop a perturbative theory accounting for the interactions between the
holes.

\section{Effective interaction between $\mathbf{t}_\mathbf{2g}$ holes: 
single $\mathbf{a}_\mathbf{1g}$ band case}
\label{sec:HtP}

\begin{figure}[tb]
\centerline{\includegraphics[width=11cm]{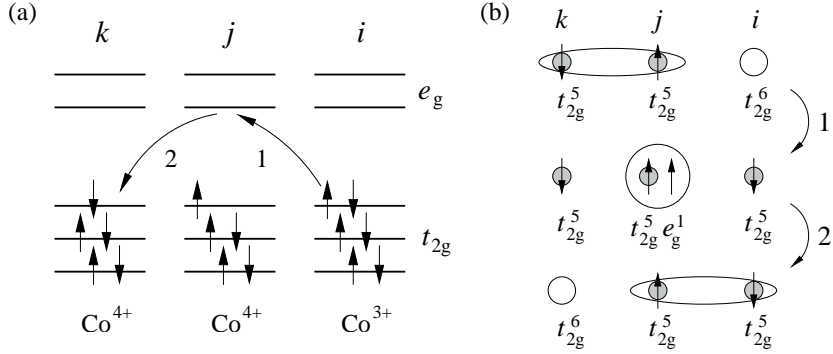}}
\caption{(a) Virtual process leading to the $H_P$ Hamiltonian in
Eq.~\eqref{eq:Heff}. $t_{2g}$ electron of a $\mathrm{Co}^{3+}_i$ ion moves to
$e_g$ level of the nearest-neighbor $\mathrm{Co}^{4+}_j$ ion (process 1) and
then to the $t_{2g}$ level of the next $\mathrm{Co}^{4+}_k$ neighbor (process
2). This is depicted in (b) as a motion of the hole-pair. The intermediate
state contains the low-lying $S=1$ state of Co$^{3+}$ ion. 
(After Ref.~\citen{Kha08}.)}
\label{fig:tPproc}
\end{figure}

In the Fermi liquid regime, the Eliashberg-formalism, where the phonon
shake-up processes (triggered by $\cal T$-exciton) can also be incorporated,
would be the best strategy. However, there are delicate constraints to handle:
a lattice site cannot be occupied by two holes or by a hole and ${\cal
T}$-excitation simultaneously. For the sake of simplicity, we derive an
effective fermionic interaction in a second order perturbation theory in
$\tilde t$ by considering the local virtual process depicted in
Fig.~\ref{fig:tPproc}. This way, all the constraints in the intermediate
states are treated explicitly. Such a perturbative treatment is valid as long
as a polaron binding energy $E_b$ (an energy gain due to the $\tilde t$
process) is small compared to a bare bandwidth $W$ ($\simeq 9t$ in a
triangular lattice). From a self-consistent Born approximation discussed
above, we obtained single polaron binding energy $E_b\simeq 2.6t$ ($2.9t$) for
$E_T$ set to the value $3t$ ($2t$) used in this paper. Hence, $E_b \sim 0.3W$
and we can integrate out virtual spin states perturbatively.

In the virtual process of Fig.~\ref{fig:tPproc} we consider an initial
configuration with a pair of Co$^{4+}$ holes and a $S=0$ Co$^{3+}$ state at
the neighboring site. First $\tilde{t}$-hopping (\textit{i.e.} the action of
$H_{\tilde{t}}$) generates an excited $S=1$ state surrounded by two Co$^{4+}$
holes. The energy of this virtual state with respect to the initial state
equals $E_T$. The virtual state gets deexcited by second $\tilde{t}$-hopping
leaving the system again with a pair of holes and a neighboring Co$^{3+}$
state. Since the pair now has a changed position, the virtual process
effectively corresponds to the motion of the hole pair. The kinetic energy
gain associated with this pair motion may lead to the pairing instability as
shown in Sec.~\ref{sec:SCsimple}. Summing up the contributions of all
possible virtual states, we arrive at the following effective Hamiltonian:
\begin{equation}
\label{eq:Heff}
H_P=\sum_{\langle ijk\rangle} \left[
 P^S_{ijk}\hat{S}^\dagger_{ij}\hat{S}^{\phantom{\dagger}}_{kj}+
 P^T_{ijk}\hat{\vc{T}}^\dagger_{ij}
      \hat{\vc{T}}^{\phantom{\dagger}}_{kj}\right] \;.
\end{equation}
The Hamiltonian in Eq.~\eqref{eq:Heff} describes the motion of the spin-singlet 
$\hat{S}_{ij}=(f_{i\uparrow}f_{j\downarrow}-f_{i\downarrow}f_{j\uparrow})/\sqrt{2}$ 
and spin-triplet 
$\hat{\vc T}_{ij}= \{f_{i\uparrow}f_{j\uparrow},
(f_{i\uparrow}f_{j\downarrow}+f_{i\downarrow}f_{j\uparrow})/\sqrt{2},
f_{i\downarrow}f_{j\downarrow}\}$ 
Co$^{4+}$--Co$^{4+}$ pairs in a background of $S=0$ Co$^{3+}$ ions. Sites
$i\neq k$ are the nearest neighbors of site $j$. No-double-occupancy
constraint on $f$ is implied when using this effective Hamiltonian. The
pair-hopping amplitudes read as
\begin{equation}\label{eq:PST}
P^S_{ijk}=\frac12 V \cos(\phi_{ij}-\phi_{jk})\;, \quad
P^T_{ijk}=\frac13 P^S_{ijk}\;.
\end{equation}
We introduced here a parameter $V=\tilde{t}^2/E_T$ characterizing the
interaction strength. The angles $\phi \in (2\pi/3, 4\pi/3, 0)$ are selected
by the orientation of the bonds $\langle ij \rangle$ and $\langle jk \rangle$
as already explained in Sec.~\ref{sec:Htttilde}. 

\begin{figure}[b]
\centerline{\includegraphics[width=11cm]{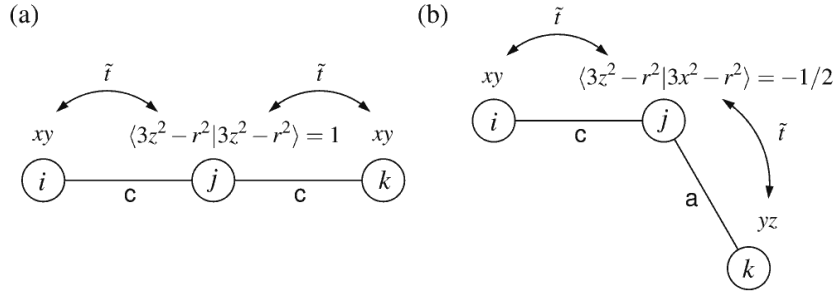}}
\caption{The origin of the geometrical factor $\cos(\phi_{ij}-\phi_{jk})$ in
Eq.~\eqref{eq:Heff} due to the overlap of the $e_g$ orbitals in the intermediate
state. (a) Along the $c$-bond, the $xy$ and $3z^2-r^2$ orbitals are coupled
via $\tilde{t}$-hopping. If the virtual process of Fig.~\ref{fig:tPproc}
proceeds along the $c$-bond exclusively, we get the geometrical factor 
$\cos(0)=1$. (b) The bond-directions for the two $\tilde{t}$-hoppings differ. 
In the case of $a$-bond, the $yz$ and $3x^2-r^2$ orbitals are coupled via
$\tilde{t}$-hopping. By taking the overlap of the $e_g$ orbitals active on the
two bonds, we get the geometrical factor $\cos(2\pi/3)=-1/2$.}
\label{fig:tPgeom}
\end{figure}

\begin{figure}[b]
\centerline{\includegraphics[width=13cm]{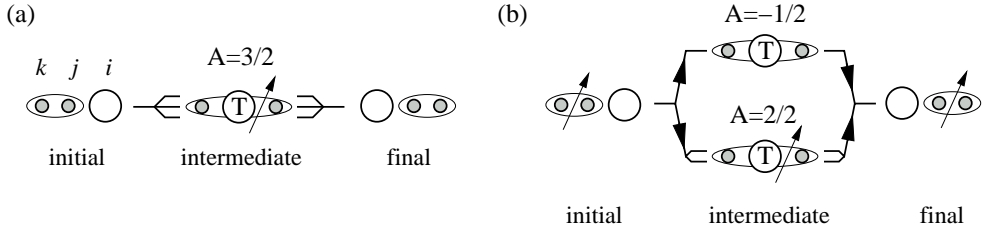}}
\caption{(a) Singlet-pair motion via an intermediate state composed of $S=1$
Co$^{3+}_j$ ion ($\mathcal{T}$ exciton) and triplet state of the two holes at
sites $i$ and $k$. The process equally uses all three $S_z=\pm 1, 0$ states of
Co$^{3+}$. The relative amplitude resulting from spin algebra equals $3/2$.
(b) The same for triplet-pair motion. Here the amplitude is distributed into
two channels -- with the two holes in the intermediate state being in singlet
(one $S_z$ state of Co$^{3+}$ employed) or triplet state (two $S_z$ states of
Co$^{3+}$ employed). Destructive interference of the two channels makes
triplet-pair hopping amplitude $3$ times smaller than that of the
singlet-pair.}
\label{fig:tPspin}
\end{figure}

Two important remarks should be made on the pair-hopping amplitudes $P_{ijk}$.
First, they contain a geometrical factor $\cos(\phi_{ij}-\phi_{jk})$ which
equals $+1$ for straight pair hopping and $-1/2$ if the hopping process
changes the direction of the pair. This factor is explained in
Fig.~\ref{fig:tPgeom}. It originates from the fact, that different $e_g$
orbitals participate in the $\tilde{t}$-hopping process on bonds of different
directions. Second, the pair hopping amplitude of singlet pairs is three
times bigger than that of the triplet pairs. This nontrivial result
originates from a quantum interference between different realizations of the
virtual process as indicated in Fig.~\ref{fig:tPspin}. The $S=1$ intermediate
state is fully transparent for singlets which equally use all three $S_z=\pm
1, 0$ states. However, these states contribute with different signs in case of
triplets, resulting in a ``spin blockade'' for their motion. 

\begin{figure}[tb]
\centerline{\includegraphics[width=9cm]{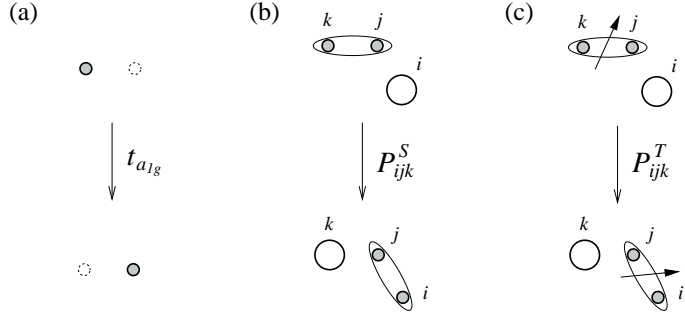}}
\caption{Cartoon representation of the processes contained in the $t-P$
Hamiltonian \eqref{eq:Ht}+\eqref{eq:Heff}: (a) $t$-hopping of $a_{1g}$ holes,
(b) pair-hopping of a singlet pair of holes, and (c) pair-hopping of a triplet
pair of holes.}
\label{fig:tPproc2}
\end{figure}

Finally, the processes contained in the reduced model Hamiltonian
$H_{t-P}=H_t+H_P$, Eqs.~\eqref{eq:Ht} and~\eqref{eq:Heff}, are summarized in a
concise way in Fig.~\ref{fig:tPproc2}. In the following sections we will use
this model containing: 1.~the usual $t$-hopping of holes
[Fig.~\ref{fig:tPproc2}(a)] and 2.~pair-hopping of hole pairs
[Fig.~\ref{fig:tPproc2}(b,c)] as the lowest-order effect of
$\tilde{t}$-hopping.

\section{Spin susceptibility}
\label{sec:spinsusc}

The pair-hopping interaction in Eq.~\eqref{eq:Heff} may be alternatively
represented in a form of density-density and spin-spin couplings, emphasizing
its relevance also to the charge and spin orderings:
\begin{equation}
\label{eq:Heff2}
H_P=V\sum_{\langle ijk\rangle} \cos(\phi_{ij}-\phi_{jk}) \left[
n_{j} n_{ik}-\tfrac13 \vc{s}_{j}\vc{s}_{ik} \right] \;.
\end{equation}
Here 
$n_j=\frac12\sum_\sigma f^\dagger_{j\sigma}f^{\phantom{\dagger}}_{j\sigma}$
and
$\vc{s}_j^z=\frac12\sum_\sigma \sigma
f^\dagger_{j\sigma}f^{\phantom{\dagger}}_{j\sigma}$
are the usual on-site charge and spin-density operators, but $n_{ik}$ and
$\vc{s}_{ik}$ with $i \neq k$ are the charge and spin densities
residing on bonds, {\it i.e.}, 
$n_{ik}=\frac12\sum_\sigma
f^\dagger_{i\sigma}f^{\phantom{\dagger}}_{k\sigma}$,
$\vc{s}_{ik}^z=\frac12\sum_\sigma \sigma
f^\dagger_{i\sigma}f^{\phantom{\dagger}}_{k\sigma}$.
Thus, the interaction acts between the local ({\it on-site}) and non-local
({\it bond}) operators which is a consequence of its three-site nature.

In momentum space, Eq.~\eqref{eq:Heff2} can be written as 
\begin{equation}
\label{eq:Heff3}
H_P= 2V\sum_{\vc{q}} \left[
n_{-\vc{q}} \tilde{n}_{\vc{q}}-\tfrac13 
\vc{s}_{-\vc{q}} \tilde{\vc{s}}_{\vc{q}}
\right]\;, 
\end{equation}
with the usual operators 
$n_{\vc{q}}=
\tfrac12 \sum_{\vc{k},\sigma}
f^\dagger_{\vc{k}+\vc{q},\sigma}
f^{\phantom{\dagger}}_{\vc{k},\sigma}$, 
$s^z_{\vc{q}}=
\tfrac12 \sum_{\vc{k},\sigma}\sigma 
f^\dagger_{\vc{k}+\vc{q},\sigma}
f^{\phantom{\dagger}}_{\vc{k},\sigma}$,
and the momentum counterparts of the non-local operators 
$\tilde{n}_{\vc{q}}=
\tfrac12 \sum_{\vc{k},\sigma}
F_{\vc{k}+\vc{q},\vc{k}}
f^\dagger_{\vc{k}+\vc{q},\sigma}
f^{\phantom{\dagger}}_{\vc{k},\sigma}$, 
$\tilde{s}^z_{\vc{q}}=
\tfrac12 \sum_{\vc{k},\sigma}\sigma 
F_{\vc{k}+\vc{q},\vc{k}}
f^\dagger_{\vc{k}+\vc{q},\sigma}
f^{\phantom{\dagger}}_{\vc{k},\sigma}$.
The formfactor $F_{\vc{k'},\vc{k}}=
\cos(k_a+k'_a) + \cos(k_b+k'_b) + \cos(k_c+k'_c)
- c_ac'_b - c_bc'_a - c_bc'_c - c_cc'_b - c_cc'_a - c_ac'_c$, 
where $c'_\alpha=\cos k'_\alpha$, originates from a peculiar bond-dependence
of interactions in Eq.~\eqref{eq:Heff2}. It manifests again that the
$\tilde{n}_q$ and $\tilde{\vc{s}}_q$ operators correspond to the
particle-hole excitations that modulate the charge and spin bonds,
respectively.

\begin{figure}[tb]
\centerline{\includegraphics[width=11.5cm]{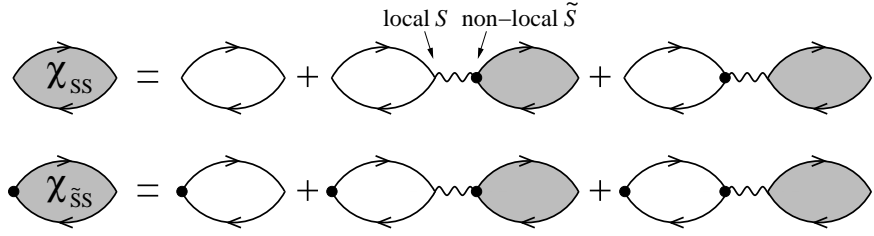}}
\caption{Diagrammatic representation of RPA equations \eqref{eq:RPA} for the
spin susceptibilities involving the interaction \eqref{eq:Heff3} between local
($\vc s_{\vc q}$) and non-local ($\tilde{\vc s}_{\vc q}$) spin densities that
reside on sites and bonds respectively.
Bare and RPA enhanced susceptibilities are represented by empty and shaded
bubbles, respectively.}
\label{fig:RPAeqs}
\end{figure}

To illustrate this unusual, nonlocal nature of correlations we investigate the
effect of the interaction on the spin susceptibility within the RPA
approximation. The bare spin susceptibility $\chi_{ss}$ is given by the formula
\begin{equation}\label{eq:suscbare}
\chi_{ss}^{(0)}(\vc q,\omega)= 
\frac1{4}\sum_{\vc k}
\frac{\tanh(\xi_{\vc k}/2T)-\tanh(\xi_{\vc k+\vc q}/2T)}
{\omega+\xi_{\vc k}-\xi_{\vc k+\vc q}+\im\delta} \;.
\end{equation}
A diagrammatic representation of the RPA approximation in our case of
local-nonlocal spin interaction is shown in Fig.~\ref{fig:RPAeqs}. RPA
enhanced susceptibility follows from the equations
\begin{equation}\label{eq:RPA}
\begin{split}
\chi_{ss} &= \chi_{ss}^{(0)} 
  + \Lambda\, \chi_{ss}^{(0)}\chi_{\tilde{s}s}
  + \Lambda\, \chi_{s\tilde{s}}^{(0)}\chi_{ss} \;, \\
\chi_{\tilde{s}s} &= \chi_{\tilde{s}s}^{(0)}
  + \Lambda\, \chi_{\tilde{s}s}^{(0)}\chi_{\tilde{s}s}
  + \Lambda\, \chi_{\tilde{s}\tilde{s}}^{(0)}\chi_{ss} \;,
\end{split}
\end{equation}
where we defined a coupling constant $\Lambda=2V/3=2\tilde{t}^2/3E_T$.
Because of the non-local spin density involved in the interaction, 
the corresponding bare susceptibilities 
$\chi_{s\tilde{s}}^{(0)}=\chi_{\tilde{s}s}^{(0)}$ and 
$\chi_{\tilde{s}\tilde{s}}^{(0)}$ are also employed. They are obtained by
multiplying the summed terms in Eq.~\eqref{eq:suscbare} by a factor of
$F_{\vc{k}+\vc{q},\vc{k}}$ 
(for $\chi_{s\tilde{s}}^{(0)}$) or 
$F^2_{\vc{k}+\vc{q},\vc{k}}$ 
(for $\chi_{\tilde{s}\tilde{s}}^{(0)}$).
The resulting spin susceptibility reads 
\begin{equation}\label{eq:RPAsusc}
\chi_{ss} = \frac{\chi_{ss}^{(0)}}{D} \;, \quad
D={\left( 1-\Lambda \chi_{\tilde{s}s}^{(0)}\right)^2- 
\Lambda^2\chi_{ss}^{(0)}\chi_{\tilde{s}\tilde{s}}^{(0)}} \;.
\end{equation}
The no-double-occupancy constraint imposed on $H_{t-P}$ handled on a
Gutzwiller level leads to a rescaling $\bar{\xi}=g_t \xi$ and $\bar{V}=g_PV$.
Since the interaction term comes from the hopping, the Gutzwiller factors are
equal $g_t=g_P\approx n_d$ and may be partially canceled in terms such as
$\Lambda\chi^{(0)}$ leaving us with an effectively rescaled $\omega$ and $T$.

\begin{figure}[t]
\centerline{\includegraphics[width=10cm]{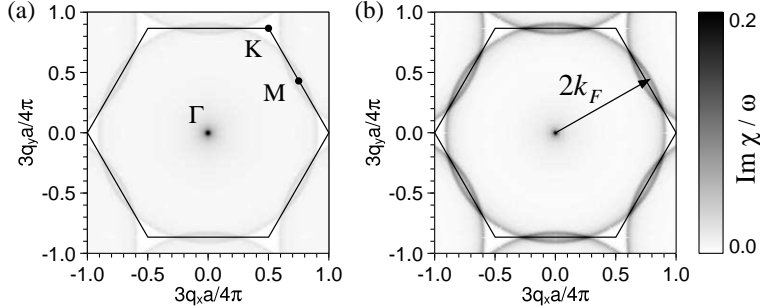}}
\caption{(a)~Map of bare $\chi_s''/\omega$ for $n_d=0.5$ (the average
Co-valency $3.5$) at $T=0.025t$ and $\omega\rightarrow 0$. (b)~Corresponding
RPA-enhanced susceptibility calculated at $\tilde{t}^2/E_T=3t$. The
interaction enhances the susceptibility at the $2k_F$ ring which (at given
density $n_d=0.5$) nearly matches the Brillouin zone boundary. (After
Ref.~\citen{Kha08}.)}
\label{fig:RPAsusc}
\end{figure}

In Fig.~\ref{fig:RPAsusc} we show a sample result at the relative fraction of
Co$^{3+}$ ions $n_d=0.5$. The bare spin susceptibility
[Fig.~\ref{fig:RPAsusc}(a)] is concentrated around the $\Gamma$ point. When
the interaction is switched on [Fig.~\ref{fig:RPAsusc}(b)], the $2k_F$ ring in
the susceptibility is enhanced. This suggests the fermionic
$2k_F$-instabilities in a {\it Fermi-liquid} phase, consistent with a picture
inferred from the experiment.\cite{Bob06} Interestingly, the RPA-spin
susceptibility at $n_d=0.5$ is most enhanced near the $M$ point, {\it i.e.}
near the observed magnetic Bragg peak position,\cite{Gas06} rather than at
$K$ typical for the AF Heisenberg spin system. In order to study the spin
ordering at $n_d=0.5$ more quantitatively, one should take into account also
the Na ordering\cite{Foo04} which breaks a hexagonal symmetry of the
underlying Fermi-surface.

The $2k_F$ antiferromagnetic correlations within our model are manifested 
in the temperature dependence of the nuclear spin relaxation rate
$T_1^{-1} \sim (\sum_{\vc{q}} 
\mathrm{Im}\,\chi/\omega)_{\omega\rightarrow 0}$.
According to the Korringa law, in the normal state $1/T_1T$ should be
temperature independent. This is the type of behavior experimentally observed
in monolayer hydrate of \cobaltate{}. However, in bilayer hydrates, which show
a superconducting transition at $T_c$, $1/T_1T$ is enhanced at low
temperatures.\cite{Kob03,Iha06} This enhancement is stronger as $T_c$ 
increases, eventually leading to a magnetic order. 

We characterize the effect of the interaction \eqref{eq:Heff2} on $1/T_1T$
by the enhancement factor $\langle F^2\rangle$
\begin{equation}
\left(\frac1{T_1 T}\right)_\mathrm{RPA}
=\langle F^2 \rangle
\left(\frac1{T_1 T}\right)_\mathrm{bare}
\end{equation}
Using RPA enhanced $\chi_{ss}$ of Eq.~\eqref{eq:RPAsusc} 
it can be approximately expressed as a double Fermi-surface average
\begin{equation}
\langle F^2 \rangle \approx
\left\langle \frac1{D^2}
\left[1-\Lambda\,\mathrm{Re}\left(\chi_{s\tilde{s}}^{(0)}
       -F_{\vc k,\vc k'}\chi_{ss}^{(0)}\right)\right]^2 
\right\rangle_{\vc k,\vc k' \in\mathrm{FS}} \;,
\end{equation}
where $D(\vc q,\omega)$, $\chi_{s\tilde{s}}^{(0)}(\vc q,\omega)$ and 
$\chi_{ss}^{(0)}(\vc q,\omega)$ are evaluated at $\vc q=\vc k'-\vc k$,
$\omega\rightarrow 0$.

\begin{figure}[t]
\centerline{\includegraphics[width=9.7cm]{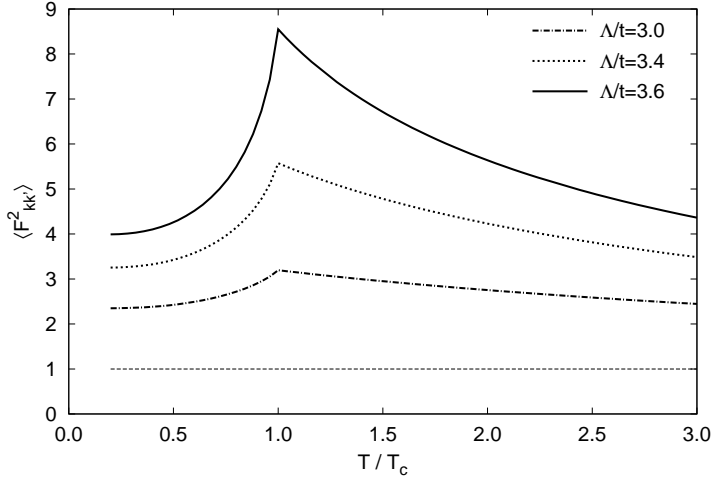}}
\caption{Enhancement factor $\langle F^2\rangle$ at $n_d=0.5$,
$t'/t=-0.2$ as a function of temperature. With the interaction switched off,
$\langle F^2\rangle=1$. The interaction enhances $1/T_1T$ near the
superconducting transition temperature as the coupling constant increases.
The critical coupling constant for a magnetic ordering in this case equals 
$\Lambda_\mathrm{crit}\approx 4 t$.}
\label{fig:RPAenhF2}
\end{figure}

The resulting enhancement factor at the relevant fraction of Co$^{3+}$
$n_d\approx 0.5$ is presented in Fig.~\ref{fig:RPAenhF2}. To address the
$T<T_c$ regime, we have included the approximate BCS gap $\Delta=1.76 T_c
\tanh (1.76\sqrt{T_c/T-1})$ in our calculations. As the interaction strength
$\Lambda$ approaches its critical value for a magnetic ordering, the
enhancement factor dramatically increases resembling the experimentally
observed critical behavior.

\section{Superconductivity due to the pair-hopping interaction}
\label{sec:SCsimple}

Now, we consider the $H_P$ Hamiltonian in the context of superconductivity.
The kinetic energy coming from the pair-hopping interaction in $H_P$ is
optimized when Co$^{4+}$ holes move pairwise. Eventually, this leads to their
condensation into SC state. The difference from cuprates is that pairs are
formed here not due to the superexchange (in cobaltates, $J$ is 
small\cite{Wan04}) but because of the gain in the kinetic energy associated with
$\tilde t$ hoppings. It is evident from Eqs.~\eqref{eq:Heff}, \eqref{eq:PST}
that spin-singlet pairs gain much more kinetic energy than triplets.
Alternatively, it can be said that the $S=0$ Co$^{3+}$ states move more
coherently when the $S=1/2$ background is in a singlet state. This result,
explained in Fig.~\ref{fig:tPspin}, is a consequence of quantum interference
among the rich variety of spin states involved during the virtual process
leading to Eq.~\eqref{eq:Heff}.

A mean-field BCS analysis\footnote{More details will be given in
Sec.~\ref{sec:SCfull} in the context of the full orbital version of our
model.} of Eq.~\eqref{eq:Heff} shows that $H_P$ supports either extended
$s$-wave singlet SC with the gap function $\propto$ $\gamma(\vc
k)=\sqrt{2/3}(c_a +c_b +c_c)$, or doubly-degenerate spin-triplet $p$-wave
pairing with $\gamma_{x,y}(\vc k)=\{(s_a-s_b), (2s_c-s_a-s_b)/\sqrt{3}\}$,
where $s_{\alpha}=\sin k_{\alpha}$. The $d$-wave channel is repulsive, while
the $f$-wave one is attractive but too weak in the physically reasonable
doping range. We estimated the $T_c$ from 
\begin{equation}\label{eq:Tceq}
1=\sum_{|\bar{\xi}_{\vc k}|\le E_T} 
\frac{\bar V_{\alpha}|\gamma_{\alpha}({\vc k})|^2}
{2\bar\xi_{\vc k}}\tanh\frac{\bar\xi_{\vc k}}{2T_c} \;,
\end{equation}
where $\bar V_{\alpha}$ is either $\bar V$ or $\bar V/3$, and the
corresponding formfactors are $\gamma(\vc k)$ or $\gamma_{x,y}(\vc k)$ for the
singlet $s$-wave and triplet $p$-wave pairing, respectively. To account for
the no-double-occupancy constraint, the fermionic dispersion as well as the
pair-hopping amplitude are renormalized by the Gutzwiller factor\cite{Zho05}
$g_t=2n_d/(1+n_d)$ as $(\bar\xi,\bar V)=(g_t\xi,g_tV)$, where $n_d$ is the
relative fraction of Co$^{3+}$ ions. In the momentum summation, we have
introduced a cutoff equal to the excitation energy $E_T$. 

\begin{figure}[tb]
\centerline{\includegraphics[width=14cm]{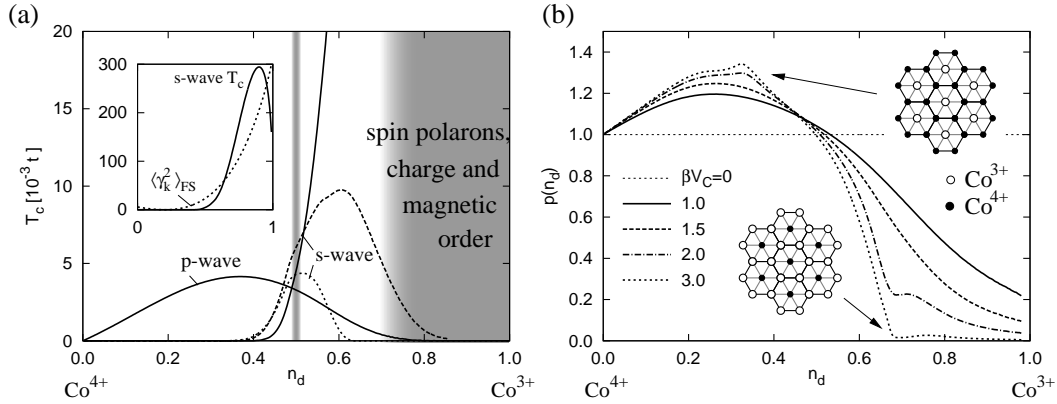}}
\caption{
(a)~$T_c$ in the extended $s$-wave and $p$-wave channels. The complete
profile of the dominant, $s$-wave $T_c$ curve is shown in the left inset
together with $\gamma_{\vc{k}}^2$ (in arbitrary units) on the Fermi
surface. The dashed ($\beta V_C=1.5$) and dotted ($\beta V_C=3$) $T_c$ curves
are calculated including nearest-neighbor Coulomb repulsion which reduces the pairing
interaction $V$ at large $n_d$. Shaded regions indicate the observed
competing orderings (including the spin-charge order at $n_d=0.5$). 
(b)~Probability ratio $p(n_d)$ (see the text for its definition) renormalizing
the pairing interaction at different values of nearest-neighbor Coulomb
repulsion relative to the effective temperature $1/\beta \propto$ bandwidth.
The feature at $n_d=1/3$ for large $V_C$ manifests a honeycomb-lattice
formation where each Co$^{3+}$ ($\circ$) has the maximum possible number of
neighboring Co$^{4+}$--Co$^{4+}$ pairs ($\bullet$--$\bullet$). Above
$n_d=2/3$, Co$^{4+}$ holes can avoid each-other completely if $V_C$ is
sufficiently large. (After Ref.~\citen{Kha08}.)}
\label{fig:Tc}
\end{figure}

We solved Eq.~\eqref{eq:Tceq} at $V=3t$ (corresponding to $\tilde t=E_T=3t$). In
terms of the BCS-coupling constant, this translates into $\lambda=\bar V\bar
N=VN \sim 1/3$ considering the density of states $N\sim 1/W\sim 1/9t$.
Therefore, the present formulation in terms of an effective fermionic
Hamiltonian \eqref{eq:Heff} should give reasonable results. At larger values
of $V$, we encounter a strong coupling regime where one should use the
original model \eqref{eq:Htilde} instead and treat virtual spin states
explicitly. This limit remains a challenging problem for future study. 

The resulting $T_c$ values from Eq.~\eqref{eq:Tceq} are presented in
Fig.~\ref{fig:Tc}(a) as solid lines. As expected, the highest $T_c$ values are
found in the singlet channel, increasing with Co$^{3+}$ density due to the
formfactor effect, until SC disappears at $n_d=1$ limit. A weak triplet
pairing is present thanks to its formfactor matching well the Fermi surface,
but it is expected to be destroyed by ({\it e.g.} Na) disorder. (We should
notice that these trends are based on the present mean-field decoupling which
ignores collective spin fluctuations. One can speculate, for instance, that
the triplet pairing may be supported by ferromagnetic fluctuations within the
CoO$_2$ planes observed\cite{Bay05} at {\it large} $n_d$ limit). 

As the SC pairing considered here is due to the pair-hopping, Coulomb
repulsion between the holes will oppose it. This is not a big trouble at high
density of Co$^{4+}$ spins (as they cannot avoid themselves) but becomes a
severe issue in a spin-diluted regime at large $n_d$, where Coulomb repulsion
reduces the process described in Fig.~\ref{fig:tPproc} hence the amplitude
$V$. Instead, the formation of spatially separated spin-polarons
(Fig.~\ref{fig:spect}) is favored, and competing orderings take over, such as
an in-plane ferromagnetism induced by a residual interactions between
spin-polarons.\cite{Kha05,Dag06} To include the effects related to the
Coulomb repulsion in the Gutzwiller fashion, we use an additional
multiplicative factor reflecting the suppression of the probability $p_{ijk}$
of having the required Co$^{3+}_i-$Co$^{4+}_j-$Co$^{4+}_k$ configuration. We
have determined this probability using a classical Monte-Carlo simulation of
hardcore particles with nearest-neighbor Coulomb repulsion $V_C$ moving on a
hexagonal lattice of $1024$ sites. The simulations were performed at
different ``effective temperatures'' $1/\beta$ imitating the kinetic energy
(of the order of bandwidth) which competes with the Coulomb repulsion in the
real system. Plotted in Fig.~\ref{fig:Tc}(b) is the probability ratio
$p(n_d)=p_{ijk}(V_C)/p_{ijk}(V_C=0)$ for several values of $\beta V_C$. The
corresponding $T_c$ curves calculated with $\bar V\rightarrow p(n_d)\bar V$
locate the SC-dome near the valence $3.4$, in a remarkable correspondence with
experiment.\cite{Tak04,Mil04,Kar04} (The reported Co-valences $\sim$3.4,
\cite{Tak04} $\sim$3.3, \cite{Mil04} $\sim$3.46 \cite{Kar04} optimal for SC
translate into $n_d=0.6,0.7,0.54$).

Finally, our $t-\tilde t$ model provides a clear hint on the role of
water-intercalation needed for SC in \cobaltate{}. Without water, a random
Na-potential induces some amount of spin-polarons locally (the origin of
``Curie-Weiss metal'' behavior\cite{Foo04}) which suppress the pairing among
the remaining fermions the usual way. Once this potential is screened-out by
the water layers, an intrinsic ground state of CoO$_2$ planes as in
Fig.~\ref{fig:Tc} is revealed. (This interpretation of the water effect is
consistent with the absence of superconductivity in the monolayer hydrate of
Na$_x$CoO$_2$, where the water resides in the Na layers.) The remaining
``enemy'' of SC is the Coulomb repulsion which prevents the pairing of dilute
Co$^{4+}$ fermions and supports the formation of spin-polarons and magnetism
instead.\cite{Kha05} More pronounced polaron physics (because of the presence
of large $S=1$ $\cal T$-exciton and narrow bandwidth) explains why $T_c$ in
cobaltates is low compared to cuprates. Another mechanism for the water effect
is provided by the band-structure calculations\cite{Joh04} that indicate a
substantial flattening of the $a_{1g}$ band-top and a reduction of the band
splitting when the water-layers are present. To study the former effect, we
have included negative $t'$ in our calculation. Due to the combined effect of
better formfactor utilization in the $s$-wave channel and Fermi velocity
reduction this enhances singlet pairing as demonstrated in Ref.~\citen{Kha08}.

\section{Full orbital structure of the interaction and 
the role of $e'_g$ pockets}
\label{sec:SCfull}

Local density approximation (LDA) bandstructure calculations on \cobaltate{}
predict the existence of $e'_g$ parts of the Fermi surface.\cite{Sin00,Lee04}
According to some theoretical explanations of the pairing mechanism in
\cobaltate{}, these so called $e'_g$ pockets may even play a crucial
role.\cite{Joh04a,Kur04,Moc05} In the ARPES experiments, however, no $e'_g$
pockets are observed in the Fermi surface, and the corresponding band of
$e'_g$ symmetry is well below the Fermi level in \cobaltate{} at the relevant
dopings. The reasons for such a discrepancy between LDA predictions and ARPES
experiments became a highly debated topic and are not yet fully 
understood.\cite{Pil08} Nevertheless, the $e'_g$ band is reported to be located
close to the Fermi level in the water-intercalated superconducting
\cobaltate{},\cite{Shi06} which opens the question of its possible role in
superconductivity. 

The aim of this section is to investigate the implications of the proximity of
$e'_g$ states to the Fermi level within our model. To this end, we first
derive a generalization of the effective interaction Hamiltonian
\eqref{eq:Heff} considering the full orbital structure of the initial,
intermediate, and final states. Using such a generalized interaction
Hamiltonian, we can study the case of mixed $a_{1g}$/$e'_g$ topmost band to
assess the effect of its $e'_g$ parts. As the model bandstructure, we use the
one from Ref.~\citen{Zho05}, where the multiorbital Gutzwiller approximation
is applied to a tight-binding fit to LDA bands. The main advantage of this
approach in the context of our study is the possibility to easily shift the
position of $e'_g$ band with respect to the Fermi level by tuning a single
tight-binding parameter.

Instead of the projected $a_{1g}$ holes we work here with the holes in canonical
$t_{2g}$ orbitals $d_{xy}$, $d_{yz}$, and $d_{zx}$. It is convenient to use
the orbital angles introduced in Fig.~\ref{fig:hopgeom}(c) to specify the
orbitals, \textit{e.g.}, $(\phi_a)_i$ annihilates a hole in $d_{yz}$ orbital at
\mbox{site $i$}. The $a_{1g}$ hole operator $f$ used in the previous sections
corresponds to the linear combination
$f_{j\sigma}=[(\phi_a)_{j\sigma}+(\phi_b)_{j\sigma}+(\phi_c)_{j\sigma}]/\sqrt{3}$
in the new notation.

\begin{figure}[tb]
\centerline{\includegraphics[width=12cm]{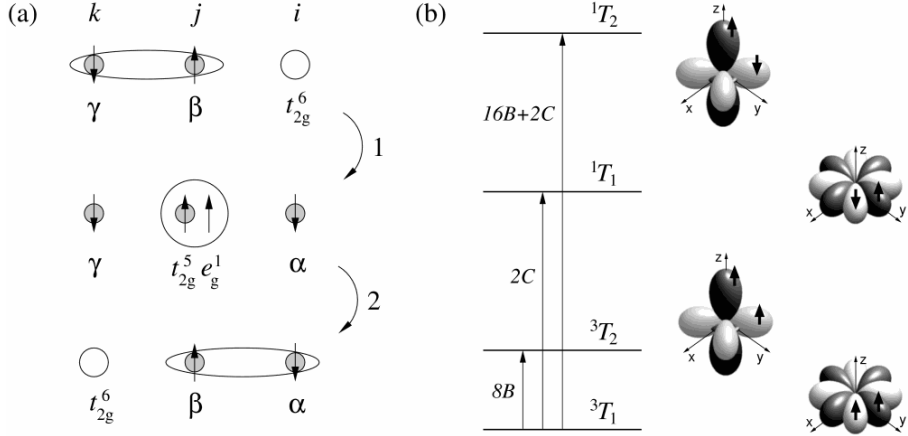}}
\caption{(a) Virtual process leading to the effective interaction between the
holes. Occupied $t_{2g}$ hole orbitals are labeled by $\alpha$, $\beta$, and
$\gamma$ respectively. The $\tilde t$-active orbitals $\alpha$ and $\gamma$
are selected by the orientation of the $\langle ij\rangle$ and $\langle jk
\rangle$ bonds. (b) Multiplet structure of a Co$^{3+}$ ion in $t_{2g}^5e_g^1$
configuration. Two lower triplet states $^3T_1$ and $^3T_2$ differ in the
occupied $e_g$ orbital being in-plane or out-of-plane with respect to the
orbital occupied by the $t_{2g}$ hole. $B$ and $C$ denote the Racah
parameters.\cite{Gri61}}
\label{fig:tPorb}
\end{figure}

To derive an effective Hamiltonian of the form \eqref{eq:Heff}, we
consider the virtual process depicted in Fig.~\ref{fig:tPorb}(a) which is an
$a_{1g}$-unprojected equivalent of the process in Fig.~\ref{fig:tPproc}.
The initial state $|i\rangle$ of the two holes at sites $k$ and $j$ is either 
singlet $S_{kj}(\gamma\beta)=(\gamma_{k\uparrow}\beta_{j\downarrow}
-\gamma_{k\downarrow}\beta_{j\uparrow})/\sqrt{2}$
or one of the triplet states
$\vc{T}_{kj}(\gamma\beta)=\{
\gamma_{k\uparrow}\beta_{j\uparrow}, 
(\gamma_{k\uparrow}\beta_{j\downarrow}
+\gamma_{k\downarrow}\beta_{j\uparrow})/\sqrt{2},
\gamma_{k\downarrow}\beta_{j\downarrow} \}$. 
The $t_{2g}$ orbitals $\gamma$ and $\beta$ of the two holes are now explicitly
indicated. The $\gamma$ orbital is the $\tilde t$-active $t_{2g}$ orbital on
$\langle jk\rangle$ bond. The amplitude of the transition to the final state
$|f\rangle$ [either $S_{ij}(\alpha\beta)$ or one of $\vc T_{ij}(\alpha\beta)$]
contributing to the effective Hamiltonian $\mathcal{H}_P$ is evaluated as
$
\langle f | \mathcal{H}_P | i \rangle = 
-\sum_{\mathrm{virt}} 
\langle f|\mathcal{H}_{\tilde t}|\mathrm{virt}\rangle
\langle\mathrm{virt}|\mathcal{H}_{\tilde t}| i\rangle / E_\mathrm{virt}\;,
$
where the summation runs over all possible intermediate states. These consist
of a hole pair ($\alpha$ hole at site $i$ and $\gamma$ hole at site $k$) and
the excited Co$^{3+}$ ion at site $j$ having the $t_{2g}^5 e_g^1$
configuration. The excitation energy depends on the spin and orbital
combination of the $t_{2g}$ hole and $e_g$ electron of the Co$^{3+}$ according
to the multiplet structure presented in Fig.~\ref{fig:tPorb}(b). The
$\tilde{t}$-hopping Hamiltonian $\mathcal{H}_{\tilde t}$ used here provides
the hopping between $t_{2g}$ and $e_g$ levels on nearest-neighbor Co ions as
in Fig.~\ref{fig:hopgeom}(b,c). The resulting $\mathcal{H}_P$ Hamiltonian
with full orbital structure can be written as
\begin{equation}\label{eq:Hefforb}
\begin{split}
{\cal H}_P=\sum_{\langle ijk\rangle}\sum_\beta ~\tilde{t}^{\,2} 
\Biggl[
 &\cos(\alpha-\beta)\cos(\gamma-\beta)
 \left(\frac32\frac1{E_{^3T_2}}-\frac12\frac1{E_{^1T_2}}\right) \\
 &+\sin(\alpha-\beta)\sin(\gamma-\beta)
 \left(\frac32\frac1{E_{^3T_1}}-\frac12\frac1{E_{^1T_1}}\right)
\Biggr] 
\;\hat{S}^\dagger_{ij}(\alpha\beta)\hat{S}^{\phantom{\dagger}}_{kj}(\gamma\beta) 
\\
+\tilde{t}^{\,2}
\Biggl[
 &\cos(\alpha-\beta)\cos(\gamma-\beta)
 \left(\frac12\frac1{E_{^3T_2}}+\frac12\frac1{E_{^1T_2}}\right) \\
 &+\sin(\alpha-\beta)\sin(\gamma-\beta)
 \left(\frac12\frac1{E_{^3T_1}}+\frac12\frac1{E_{^1T_1}}\right)
\Biggr] 
\;\hat{\vc{T}}^\dagger_{ij}(\alpha\beta)
  \hat{\vc{T}}^{\phantom{\dagger}}_{kj}(\gamma\beta) \;,
\end{split}
\end{equation}
where $\alpha$ and $\gamma$ are selected by the directions of the bonds
$\langle ij\rangle$ and $\langle jk\rangle$ respectively. The cosine and sine
factors come from the overlap of the $\tilde t$-active $e_g$ orbitals at the
respective bonds and the $e_g$ orbitals participating in the virtual states.
In the following, we neglect the contribution of the high-energy $S=0$ states
of Co$^{3+}$. The simpler version of the Hamiltonian in Eq.~\eqref{eq:Hefforb}
as given by Eq.~\eqref{eq:Heff} is obtained by letting
$E_{^3T_1}=E_{^3T_2}=E_T$ and projecting on the $a_{1g}$ states.

Next we study the pairing interaction contained in \eqref{eq:Hefforb}. In
Sec.~\ref{sec:SCsimple}, it was shown, that the singlet pairing channel leads
to the SC dome similar to the experimentally observed one. We therefore
concentrate on the singlet-hopping term in \eqref{eq:Hefforb} and estimate
$T_c$ along the same lines as in Sec.~\ref{sec:SCsimple}. The corresponding
singlet operators are first projected onto the topmost band (the closest to
$\mu$) $f_{\vc k\sigma}=u_{\vc k a} (\phi_a)_{\vc k\sigma}+ u_{\vc k b}
(\phi_b)_{\vc k\sigma}+u_{\vc k c}(\phi_c)_{\vc k\sigma}$ as the relevant one
for a $T_c$ estimation. A BCS Hamiltonian
\begin{equation}
H_\mathrm{BCS} = \sum_{\vc k \vc k'} V_{\vc k\vc k'}
f^\dagger_{\vc k\uparrow}f^\dagger_{-\vc k\downarrow}
f^{\phantom{\dagger}}_{-\vc k'\downarrow}
f^{\phantom{\dagger}}_{\vc k'\uparrow}
\end{equation}
is then extracted from the result. The formfactor $V_{\vc k\vc k'}$ has a
complicated form containing the bandstructure coefficients $u_{\vc k\alpha}$
as well as the overlap factors in a triple sum over orbital angles
\begin{multline}\label{eq:formfacorb}
V_{\vc k\vc k'}=6\tilde{t}^2
\sum_{\alpha\beta\gamma} (2-\delta_{\alpha\gamma})
u_{\vc k\alpha}u_{\vc k\beta}u_{\vc k'\beta}u_{\vc k'\gamma}
\cos k_\alpha \cos k'_\gamma \times \\
\times \left[
\frac1{E_{^3T_1}}
\sin(\alpha-\beta)\sin(\gamma-\beta) 
 +
\frac1{E_{^3T_2}}
\cos(\alpha-\beta)\cos(\gamma-\beta) 
\right] \;.
\end{multline}
In principle, Eq.~\eqref{eq:formfacorb} in its present form can be used for
the calculation of $T_c$, but to get a general insight to the results, a
symmetry analysis of this formfactor is necessary. Guided by the group theory,
we express Eq.~\eqref{eq:formfacorb} using symmetry-adapted functions.

\begin{table}[t]
\caption{Character table for the $C_{6v}$ group and a basis of its irreducible
representations derived from $\cos k_\alpha$, $\sin k_\alpha$.  These
functions are orthonormal on the BZ, \textit{i.e.}, 
$\sum_{\vc k} \gamma_i(\vc k)\gamma_j(\vc k) = \delta_{ij}$~.}
\label{tab:irr}
\begin{center}
\begin{tabular}{cccccccl} \hline\hline
representation & $E$ & $2C_6$ & $2C_3$ & 
$C_2$ & $3\sigma_v$ & $3\sigma_d$ & function \\ \hline
\rule{0mm}{7mm}
$A_1$ &1& 1& 1& 1&1& 1& 
 $\gamma_s(\vc k)=\sqrt{\frac23}\,(\cos k_a+\cos k_b+\cos k_c)$ \\
\rule{0mm}{6mm}
$B_1$ &1&-1& 1&-1&1&-1& 
 $\gamma_f(\vc k)=\sqrt{\frac23}\,(\sin k_a+\sin k_b+\sin k_c)$ \\
\rule{0mm}{6mm}
$E_1$ &2& 1&-1&-2&0& 0& 
 $\gamma_{p1}(\vc k)=\sin k_a-\sin k_b$ \\
\rule{0mm}{6mm}
 & & & & & & & 
 $\gamma_{p2}(\vc k)=\frac1{\sqrt 3}\, (2\sin k_c-\sin k_a-\sin k_b)$ \\
\rule{0mm}{6mm}
$E_2$ &2&-1&-1& 2&0& 0& 
 $\gamma_{d1}(\vc k)=\cos k_a-\cos k_b$ \\
\rule{0mm}{6mm}
 & & & & & & & 
 $\gamma_{d2}(\vc k)=\frac1{\sqrt 3}\, (2\cos k_c-\cos k_a-\cos k_b)$ \\
 & & & & & & & \\
\hline
\end{tabular}
\end{center}
\end{table}

In Table~\ref{tab:irr}, we show the irreducible representations of the $C_{6v}$
group being the symmetry group of the hexagonal lattice of Co ions. The
factors $\cos k_\alpha$ and $\cos k'_\gamma$ as well as the bandstructure
coefficients $u$ can be expressed in terms of functions belonging to $A_1$ and
$E_2$ representation of the $C_{6v}$ group ($B_1$ and $E_1$ is employed in the
case of triplet pairing). The functions $\gamma_s$, $\gamma_{d1}$, and
$\gamma_{d2}$ allow us to directly express $\cos k_\alpha$ and $\cos
k'_\gamma$, the bandstructure coefficients $u$ are converted to a
symmetry-adapted form via the relations
$a_{1g}=(u_a+u_b+u_c)/\sqrt3$,
$e'_{g1}=(u_a-u_b)/\sqrt2$,
$e'_{g2}=(2u_c-u_a-u_b)/\sqrt6$ 
deduced from Table~\ref{tab:irr}. We then perform the summations over
$\alpha$, $\beta$, and $\gamma$ and regroup\footnote{This procedure is quite
involved, since Eq.~\eqref{eq:formfacorb} at this stage contains a huge number
of terms.} the resulting terms to form products of symmetric functions of $\vc
k$ and $\vc k'$ respectively. For example, for the $A_1$ representation
(corresponding to the extended $s$-wave symmetry) there exist four such
functions
$\Psi_{s1}=\gamma_s a_{1g} a_{1g}$,
$\Psi_{s2}=\frac1{\sqrt2}\left( 
 \gamma_{d1} a_{1g} e'_{g1}+\gamma_{d2} a_{1g} e'_{g2} \right)$,
$\Psi_{s3}=\frac1{\sqrt2}\gamma_s\left(
 e'_{g1}e'_{g1}+e'_{g2}e'_{g2} \right)$,
$\Psi_{s4}=\gamma_{d1} e'_{g1}e'_{g_2}+
\frac12\gamma_{d2}({e'}_{g1}^2-{e'}_{g2}^2)$.

\begin{figure}[tb]
\centerline{\includegraphics[width=14.0cm]{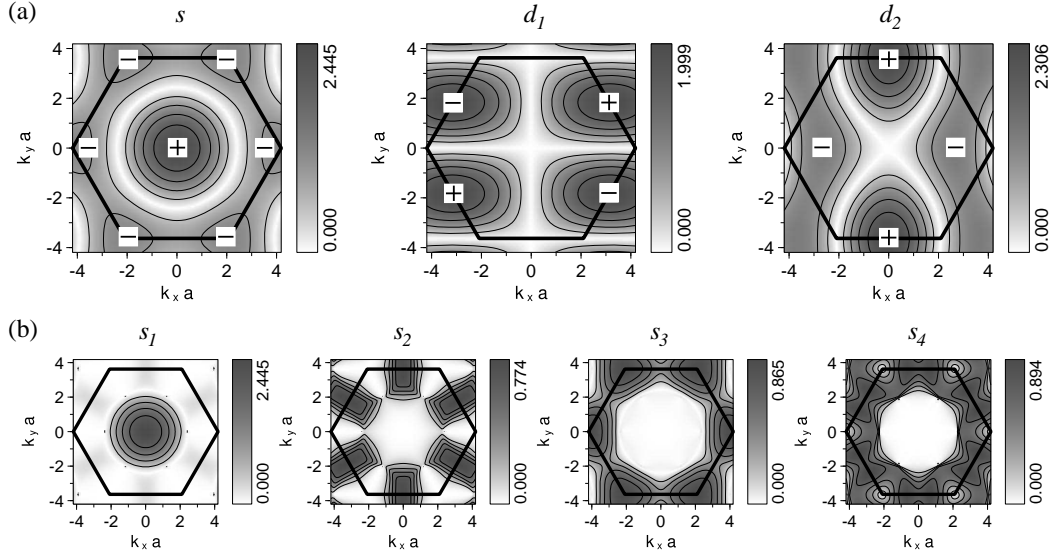}}
\caption{(a) Symmetry functions $\gamma_s(\vc k)$, $\gamma_{d1}(\vc k)$ and
$\gamma_{d2}(\vc k)$ entering the singlet-channel BCS interaction
\eqref{eq:sworbff2} resulting from \eqref{eq:Hefforb} in the simple $a_{1g}$
case. (b) Symmetry adapted functions entering the BCS interaction in the
singlet, extended $s$-wave channel in the general case. The bandstructure from
Ref.~\citen{Zho05} at $n_d=0.5$ was used here.}
\label{fig:symfun}
\end{figure}

Disentangled interaction formfactor splits into the sum of $s$-wave,
$d_1$-wave and $d_2$-wave contributions, \textit{i.e.}, these channels are
independent in our BCS Hamiltonian. The formfactor of the $s$-wave channel
reads as 
\begin{equation}\label{eq:sworbff}
V_{\vc k\vc k'}^\mathrm{(s-wave)}= 
\lambda \vc{\Psi}^T(\vc k) (M_1+r M_2) \vc{\Psi}(\vc k')
\end{equation}
where we have introduced the vector
$\vc{\Psi}^T(\vc k)=\left[\Psi_{s1}(\vc k),\Psi_{s2}(\vc k),
\Psi_{s3}(\vc k),\Psi_{s4}(\vc k)\right]$,
the coupling constant $\lambda={\tilde{t}^2}/{E_{^3T_1}}$, the ratio of the
triplet excitation energies $r=E_{^3T_1}/E_{^3T_2}$ and the matrices
$$
M_1=\frac14 \begin{pmatrix}
     -2 & -\sqrt2 &+\sqrt2 & -\sqrt2 \\
-\sqrt2 & -1      &+1      & -1      \\
+\sqrt2 & +1      & -1     &+1       \\
-\sqrt2 & -1      &+1      & -1 
\end{pmatrix},\;
M_2=\frac14 \begin{pmatrix}
      -2 & -3\sqrt2 & -\sqrt2 &+\sqrt2 \\
-3\sqrt2 & +1      &+7      & -7      \\
-\sqrt2  & +7      &+9      & -9      \\
+\sqrt2  & -7      & -9     &+9
\end{pmatrix}.
$$
Each of the two $d$-wave channels could be represented in a similar way using
$6\times6$ matrices and corresponding six symmetry adapted functions. It is
instructive to see, how a simple $a_{1g}$ expression emerges from the general
result. In such case $a_{1g}=1$ and $e'_{g1}=e'_{g2}=0$ everywhere in the
Brillouin zone implying $\Psi_{s1}=\gamma_s$,
$\Psi_{s2}=\Psi_{s3}=\Psi_{s4}=0$ and a similar reduction in the $d$-wave
part. The simplified expression for the singlet-pairing formfactor could be
written as
\begin{equation}\label{eq:sworbff2}
V_{\vc k\vc k'}=\frac{1+r}2 \lambda \left[
-\gamma_s(\vc k)\gamma_s(\vc k') 
+2\gamma_{d1}(\vc k)\gamma_{d1}(\vc k')
+2\gamma_{d2}(\vc k)\gamma_{d2}(\vc k')
\right] 
\end{equation}
which shows an attractive extended $s$-wave part and a repulsive $d$-wave
part. The symmetry functions $\gamma_s$, $\gamma_{d1}$, and $\gamma_{d2}$ are
presented in Fig.~\ref{fig:symfun} along with the functions $\Psi_{s1}$,
$\Psi_{s2}$, $\Psi_{s3}$, and $\Psi_{s4}$ for the bandstructure of
Ref.~\citen{Zho05}.

Finally, we estimate $T_c$ in the singlet, extended $s$-wave channel using
the disentangled formfactor \eqref{eq:sworbff}. Plugging in our form of
$V_{\vc k\vc k'}$ into the gap equation at $T=T_c$, we find that the
transition temperature $T_c$ is determined by the condition of the largest
eigenvalue of the $4\times4$ matrix
\begin{equation}\label{eq:Tceqorb}
-g_t \lambda (M_1+rM_2) \sum_{\vc k} 
{\vc \Psi}(\vc k) {\vc \Psi}^T(\vc k) 
\frac{\tanh (\bar{\xi}_{\vc k} / 2T_c)}{2\bar\xi_{\vc k}}
\end{equation}
being equal to $1$. Here $\bar{\xi}_{\vc k}$ is the dispersion of the 
renormalized topmost band.

\begin{figure}[tb]
\centerline{\includegraphics[width=14cm]{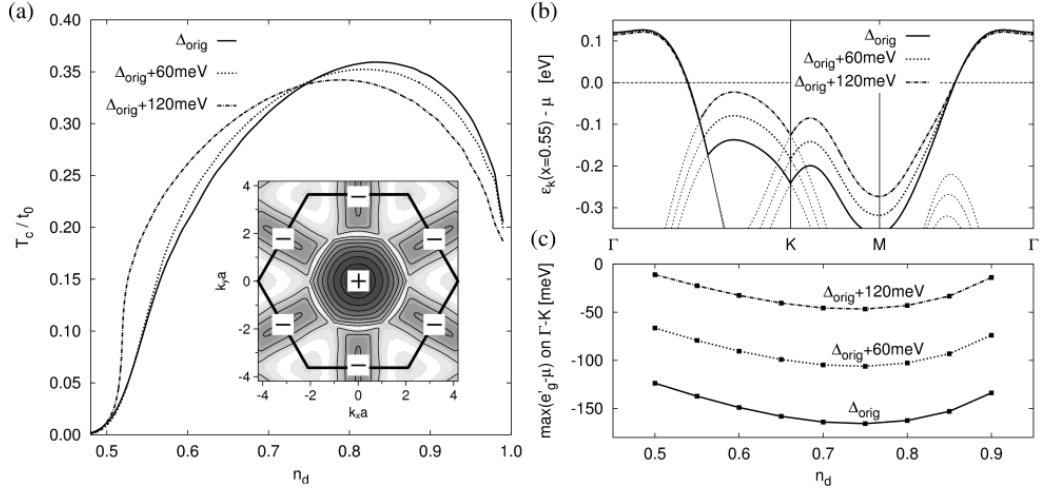}}
\caption{(a) Transition temperature $T_c$ as a function of $n_d$ for the
bandstructure of Ref.~\citen{Zho05}. The $a_{1g}$--$e'_g$ orbital splitting
$\Delta$ was modified to shift the $e'_g$ bands closer to the Fermi level.
$E_{^3T_1}=\tilde t/2$, $r=1/3$ was used in the calculation. The effect of
the nearest-neighbor Coulomb repulsion is not included. The inset shows the
gap function in the Brillouin zone. (b) The effect of the modified
$a_{1g}$--$e'_g$ orbital splitting on the bandstructure at $n_d=0.55$. Topmost
band employed in $T_c$ calculation is drawn as a solid line. (c) Distance of
the top of the $e'_g$ band to the Fermi level as a function of $n_d$.}
\label{fig:Tcorb}
\end{figure}

Shown in Fig.~\ref{fig:Tcorb} is the doping-dependent transition temperature
as found from \eqref{eq:Tceqorb} employing the topmost band of the
bandstructure of Ref.~\citen{Zho05}. To study the effect of the proximity of
$e'_g$ band to the Fermi level we have varied the $a_{1g}$--$e'_g$ orbital
splitting determining the position of the $e'_g$ band. As observed in
Fig.~\ref{fig:Tcorb}, $T_c$ in the relevant range increases as the $e'_g$ band
approaches $\mu$ very closely, to around $10-30\:\mathrm{meV}$. 

Another effect of the $e'_g$ band is a strong anisotropy of the
superconducting gap. In the pure $a_{1g}$ case, the gap function as well as
the nearest neighbor tight-binding dispersion is proportional to $\gamma_s(\vc
k)$, so that the gap is totally isotropic at the Fermi surface. In the
present case, the gap consists of the four symmetry functions presented in
Fig.~\ref{fig:symfun}(b). Near the Brillouin zone center, where the topmost
band is of $a_{1g}$ character, the gap is again determined by $\gamma_s(\vc
k)$.  However, closer to the Brillouin zone boundary, the band switches to
mainly $e'_g$ character and the functions $\Psi_{s2}$, $\Psi_{s3}$ and
$\Psi_{s4}$ start to dominate the gap reversing its sign. They bring strong
anisotropy to the gap function at the Fermi surface which attains $30-60\%$ at
$n_d=0.55$ depending on the $a_{1g}$--$e'_g$ splitting. Such a strong
anisotropy of the gap may lie behind the absence of the coherence peak in the
nuclear spin relaxation rate data.\cite{Fuj04,Ish04}

\section{Conclusions}
\label{sec:conclusions}

We have discussed a strongly correlated model for \cobaltate{} which is based
on the spin-state quasidegeneracy of $S=0$ and $S=1$ states of Co$^{3+}$ and
the specific geometry of the CoO$_2$ planes in layered cobaltates. The basic
idea of the model is that $t_{2g}-e_g$ hopping of $3d$ electrons between
nearest-neighbor cobalt ions, which is enabled thanks to the $90^\circ$
geometry of \mbox{Co-O-Co} bonds, allows to employ the low-lying $S=1$
$t_{2g}^5e_g^1$ state of Co$^{3+}$. This new degree of freedom is exploited by
the Co$^{4+}$ holes as they move in the CoO$_2$ plane.

In the sodium-rich region, when the Co$^{4+}$ holes are dilute, the model
naturally explains experimentally observed strong correlations and interprets
them in terms of a spin-polaron formation. At higher concentration of the
doped holes, when the Fermi-liquid regime is established, we have derived
effective interactions between the holes and discussed their impact on spin
fluctuations. The superconductivity mediated by the spin-state fluctuations of
Co$^{3+}$ ions emerges in the model, at experimentally observed compositions.
Finally, we discussed the symmetry of the effective interaction in the context
of the possible role of $e'_g$ bands in the superconductivity.

Given the simplicity and experimentally motivated design of the model, its
success can hardly be accidental. Therefore, $t-\tilde t$ Hamiltonian,
Eqs.~\eqref{eq:Ht} and~\eqref{eq:Htilde}, can be regarded as a basic minimal
model for \cobaltate{}. The idea of spin-state fluctuations may also have
broader applications, {\it e.g.}, in oxides of Rh and Ir ions with a similar
spin-orbital structure and lattice geometry. 

Formally, systems like LaCoO$_3$ and NaCoO$_2$ fall into the category of band
insulators because Co$^{3+}$ ions have an even number of electrons forming a
completely filled, spinless configuration $t_{2g}^6$ in their ground state.
Nevertheless, they are strongly correlated materials due to a proximity of 
virtual magnetic configurations of Co$^{3+}$ which can easily be activated by
doping, temperature, {\it etc}. We conclude that the presence of such 
low-lying magnetic states is responsible for the rich and nontrivial physics
of cobaltates. 

\section*{Acknowledgments}
We would like to thank B.~Keimer for stimulating discussions. 
This work was partially supported by the Ministry of Education of Czech
Republic (MSM0021622410).

\end{document}